\documentclass[twocolumn,pra,superscriptaddress,longbibliography]{revtex4-2}
\usepackage{graphicx,amsmath,amssymb}
\usepackage[colorlinks=true, citecolor=blue, urlcolor=blue, linkcolor=blue]{hyperref}

\begin{document}
	
	\title{Reinforcement-learning-assisted nonreciprocal optomechanical 
		gyroscope}
	
	\author{Qing-Shou Tan}
	\email{qstan@hnist.edu.cn}
	\affiliation{Key Laboratory of Hunan Province on Information Photonics and
		Freespace Optical Communication, College of Physics and Electronics,
		Hunan Institute of Science and Technology, Yueyang 414000, China}
	\author{Ya-Feng Jiao}
	\affiliation{School of Electronics and Information, Zhengzhou University of 
		Light Industry, Zhengzhou 450001, China}
	\affiliation{Academy for Quantum Science and Technology, Zhengzhou 
		University of Light Industry, Zhengzhou 450002, China}
	\author{Yunlan Zuo}
	\affiliation{School of Physics and Chemistry, Hunan First Normal University, Changsha 410205, China}
	\author{Lan Xu}
	\affiliation{School of Physics and Chemistry, Hunan First Normal University, Changsha 410205, China}
	\author{Jie-Qiao Liao}
	\affiliation {Key Laboratory of Low-Dimensional Quantum Structures and 
		Quantum Control of Ministry of Education,
		Department of Physics and Synergetic Innovation Center for Quantum 
		Effects and Applications, Hunan Normal University, Changsha 410081, 
		China}
	\affiliation {Institute of Interdisciplinary Studies, Hunan Normal University, Changsha, Hunan 410081, China}
	\author{Le-Man Kuang}
	\email{lmkuang@hunnu.edu.cn}
	\affiliation {Key Laboratory of Low-Dimensional Quantum Structures and 
		Quantum Control of Ministry of Education, Department of Physics and 
		Synergetic Innovation Center for Quantum Effects and Applications,
		Hunan Normal University, Changsha 410081, China}
	\affiliation{Academy for Quantum Science and Technology, Zhengzhou 
		University of Light Industry, Zhengzhou 450002, China}
		\affiliation {Institute of Interdisciplinary Studies, Hunan Normal University, Changsha, Hunan 410081, China}
	
	\begin{abstract}
		
		We propose an optomechanical gyroscope architecture based on a spinning cavity optomechanical  resonator  evanescently coupled to a tapered optical fiber without relying on costly quantum light sources. Our study reveals a striking dependence of the gyroscope's sensitivity on the propagation direction of the driving optical field, manifesting robust quantum nonreciprocal behavior. This nonreciprocity significantly enhances the precision of angular velocity estimation, offering a unique advantage over conventional gyroscopic systems. Furthermore, we demonstrate that the operational range of this nonreciprocal gyroscope is fundamentally governed by the frequency of the pumping optical field, enabling localized sensitivity to angular velocity. Leveraging the adaptive capabilities of reinforcement learning (RL), we optimize the gyroscope's sensitivity within a targeted angular velocity range, achieving unprecedented levels of precision. These results highlight the transformative potential of RL in advancing high-resolution, miniaturized optomechanical gyroscopes, opening new avenues for next-generation inertial sensing technologies.
	\end{abstract}

	\maketitle
	\section{Introduction}
	
	   The gyroscope, an indispensable instrument for measuring angular velocity in moving objects, plays a pivotal role in navigation and inertial guidance systems~\cite{RevModPhys.57.61,PhysRevD.86.083002}. Optical gyroscopes, the most prevalent variant, determine rotation rates by detecting the phase shift between counter-propagating beams within an optical loop, leveraging the well-known Sagnac effect~\cite{RevModPhys.39.475,OL1981,kuang2025}. However, achieving high precision necessitates enlarging the optical loop's coverage area, leading to significant challenges such as increased size, elevated power consumption, and difficulties in reconciling sensitivity with integration. Recent advancements in quantum sensing have reignited interest in developing compact, high-precision quantum gyroscopes capable of surpassing the 
	   standard quantum limit (SQL)~\cite{PhysRevA.57.4736,PhysRevA.95.012326,PhysRevLett.124.120403}.

       Significant progress in optomechanics has enabled unprecedented nanoscale manipulation of light~\cite{RevModPhys.86.1391,RevModPhys.86.1391,Verhagen2012,Davuluri_2017}. Cavity optomechanical (COM) systems, which integrate optical and mechanical modes, offer a versatile platform for exploring the interface between macroscopic and microscopic realms, shedding light on fundamental quantum phenomena~\cite{PhysRevLett.107.020405, PhysRevLett.113.020405}. The advent of COM has extended the study of macroscopic quantum effects to massive mechanical oscillators. Optical microcavities, such as photonic crystal cavities and microwave superconducting cavities, provide distinct advantages including compact size, high vibrational frequency, and chip-level integration, making them highly suitable for quantum sensing applications~\cite{Massel2012,LPR2011,Massel2012}, including the development of optomechanical gyroscopes~\cite{Li:22,sun2023}.
 
       In this work, we present a compact and integrable optomechanical gyroscope based on a spinning COM resonator.
       Recent studies have demonstrated that the Sagnac effect in spinning optical microcavities can induce frequency shifts in opposite directions~\cite{Maayani2018}, thereby breaking the system’s time-reversal symmetry and enabling diverse nonreciprocal quantum  effects~\cite{PhysRevLett.125.143605, PhysRevApplied.18.064008, PhysRevLett.121.153601,  PhysRevLett.132.193602, PhysRevLett.133.043601}. These nonreciprocal quantum effects, characterized by enhanced coherence and robustness, offers appealing resources for a variety of emerging quantum technologies, ranging from quantum computing to quantum sensing. In particular, the Sagnac effect in spinning COM systems induces changes in light propagation velocity or phase~\cite{PhysRevLett.125.143605, PhysRevApplied.18.064008}, providing a potential platform for precise angular velocity measurement through optical monitoring~\cite{Jing:18, Zhang2020, Appl.Phys.Rev.11.031409}.
      The spinning COM resonator’s geometric configuration supports degenerate clockwise (CW) and counterclockwise (CCW) optical modes. 
	  Pump light injected into the optical fiber evanescently couples into the microsphere cavity, forming a resonant loop. Rotation induces Sagnac-Fizeau frequency shifts in the CW and CCW modes, altering their quantum properties and enabling precise rotation speed measurement~\cite{Maayani2018}. By accounting for system dissipation and environmental noise, we solve the master equation and input-output relations for the CW and CCW modes, deriving the quantum Fisher information (QFI) for rotation speed~\cite{Helstrom1969}. Notably, our approach surpasses the SQL without requiring expensive quantum light sources. Exploiting quantum non-reciprocity through the Sagnac effect and optimizing the pump light direction further enhance sensitivity, marking a significant advancement in quantum metrology.

	  Theoretical analysis reveals that the gyroscope’s high sensitivity is constrained by the pumping optical field’s frequency, indicating localized angular velocity sensitivity. To overcome this limitation, we introduce a novel approach leveraging reinforcement learning (RL) techniques to enhance sensitivity across a broader angular velocity range. RL, a subset of machine learning, has demonstrated remarkable success in quantum control and metrology~\cite{Xu2019,PhysRevLett.125.100401,
		Porotti2019,PhysRevLett.126.060401,PhysRevX.8.031084,Ai2022,Zhang2023,PhysRevLett.132.213602,PhysRevA.103.032601,PhysRevX.8.031086, PhysRevLett.131.050601, Xiao2022,PhysRevResearch.4.043057,PhysRevA.109.042417}. 
	  By optimizing the detuning between the pump light and spinning cavity frequencies, RL significantly improves sensitivity over an extended operational range, showcasing its transformative potential for next-generation quantum sensing devices.
	
	  The paper is structured as follows: Sec.~\ref{II} outlines the physical model of the nonreciprocal optomechanical gyroscope. Section~\ref{III} derives the QFI formula to evaluate sensitivity. Section~\ref{IV} presents numerical simulations demonstrating sensitivity enhancement through non-reciprocity. Section~\ref{V} explores RL-assisted global sensitivity optimization. Finally, Sec.~\ref{VI} provides concluding remarks.

	\begin{figure}[tp]
		\centering\includegraphics[width=8.6cm]{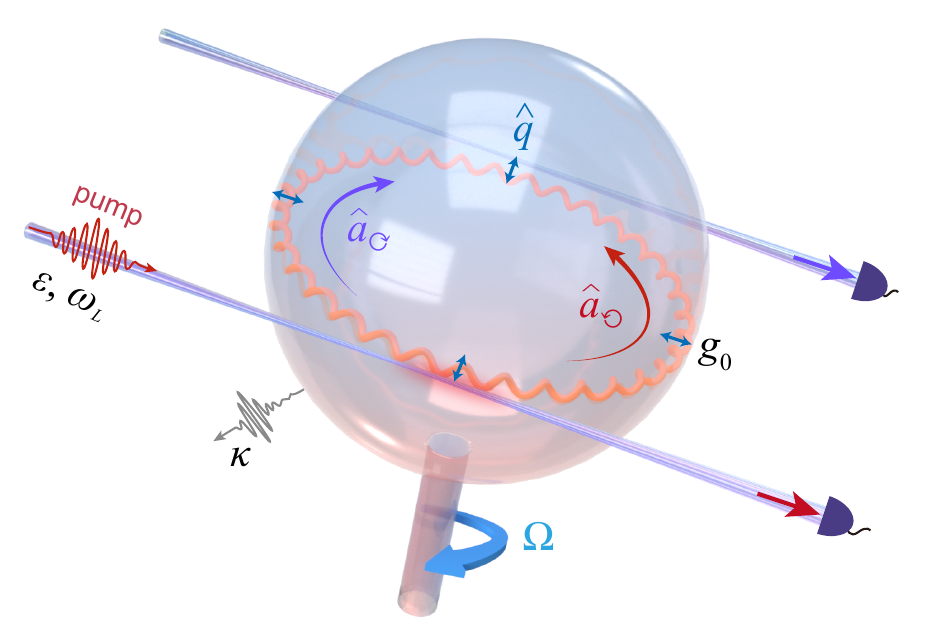}
		\caption{Schematic of a nonreciprocal optical gyroscope in a rotating optomechanical system.
			This schematic illustrates a rotating whispering-gallery-mode (WGM) microsphere cavity, where light circulates in both CW and CCW directions, forming distinct resonant modes. The input pump light is coupled into the cavity through evanescent waves, generating a standing wave as it circulates in resonance. The cavity’s mechanical breathing mode interacts with the optical field via radiation pressure, characterized by a coupling strength denoted by \(g_0\).
			Under rotation, the CW and CCW modes experience opposite frequency shifts \(\Delta_F\) due to the Sagnac effect. This non-reciprocal frequency shift forms the basis for precise and directional measurement of rotational velocity. High-precision sensing of angular velocity \(\Omega\) can be achieved by measuring the output optical field from the WGM cavity.
			 }
		\label{Fig1}
	\end{figure}

	\section{THE PHYSICAL MODEL} \label{II}
	
	We consider an optomechanical gyroscope consisting of a spinning COM resonator that is evanescently coupled to a tapered fiber. The system's rotation induces distinct optical path differences for counter-propagating modes within the resonator, leading to asymmetric refractive indices for the  CW and CCW modes. Consequently, the resonance frequencies undergo opposite Sagnac-Fizeau shifts~\cite{PhysRevLett.125.143605, Grigorii2000}, given by  
	$ \omega_c \to \omega_c + \Delta_F[\pm\Omega]$,
	where  $\Delta_F[\pm\Omega] \sim \pm \Omega n R \omega_c / c.$
	Here, $\omega_c$ is the resonance frequency of a stationary resonator, $n$ is the refractive index, $\Omega$ is the angular velocity, $R$ is the resonator radius, and $c$ is the speed of light in vacuum. When the resonator spins in the CW direction, the frequency shift satisfies $\Delta_F > 0$ for left-hand side driving and $\Delta_F < 0$ for right-hand side driving, leading to effective optical frequencies  
	$ \omega_{j= \circlearrowleft, \circlearrowright } \equiv \omega_c \pm |\Delta_F|,$
	as illustrated in Fig.~\ref{Fig1}. Additionally, the resonator supports a mechanical breathing mode with frequency $\omega_m$. The rotational angular velocity $\Omega$ significantly influences the optical field within the resonator, enabling precise measurements of $\Omega$ through the optical field characteristics.
	
	When the cavity is driven from the CCW direction, the Hamiltonian governing the rotating optomechanical system is given by~\cite{PhysRevLett.125.143605}  (with $\hbar=1$)
	\begin{equation}\label{hami}
		H   = H_0  + H_{\rm dr} + H_{\rm int},
	\end{equation}
	with  
	\begin{subequations}
		\begin{align}
			H_0 &=\sum_{j= \circlearrowleft, \circlearrowright} 
			\omega_{j}a^{\dagger}_j a_j + \omega_m b^{\dagger}b, \\
			H_{\rm dr} &= i \varepsilon ( a_{\circlearrowleft}^{\dagger} 
			e^{-i\omega_L t}- a_{\circlearrowleft}e^{i\omega_L t}), \\
			H_{\rm int} &= 
			-g_0\sum_{j= \circlearrowleft, \circlearrowright} 
			a^{\dagger}_j a_j (b^{\dagger} 
			+ b)
			+J(a_{\circlearrowright}^{\dagger}a_{\circlearrowleft} 
			+a_{\circlearrowleft}^{\dagger}a_{\circlearrowright} ).
		\end{align}
	\end{subequations}
	Here, $a_j$ ($a_j^{\dagger}$) and $b$ ($b^{\dagger}$) denote the annihilation (creation) operators of the optical cavity field and the mechanical oscillator, respectively.  
	In Eq.~(\ref{hami}), $H_0$ describes the free Hamiltonian of the system, while $H_{\rm dr}$ represents the external driving of the CCW cavity mode. The pulsed field has a carrier frequency $\omega_L$ and a pump amplitude given by $\varepsilon=\sqrt{2\kappa P/\hbar\omega_L}$, where $\kappa$ is the cavity decay rate, and $P$ is the input laser power.  
	
	The interaction Hamiltonian $H_{\rm int}$ accounts for couplings between the subsystems. Specifically, the first term describes the optomechanical interaction induced by radiation pressure, characterized by the single-photon COM coupling rate  
	$ g_0 = (\omega_c/R)\sqrt{\hbar/m\omega_m},$
	where $m$ is the mass of the resonator~\cite{RevModPhys.86.1391}. The second term represents the coupling between CW and CCW optical modes due to backscattering, which arises from surface roughness or material defects. The parameter $J$ quantifies the strength of this coupling.  
	
	In the open-system case,  the two cavity field models are assumed to be 
	coupled with vacuum fields,  and the mechanical oscillator 
	is coupled to a heat bath.
	The quantum dynamics of the coupled optomechanical system (comprising optical cavity modes and a mechanical resonator mode) are described by the following Born-Markovian master equation~\cite{breuer2002theory,PhysRevA.109.063508, PhysRevA.110.023519}:
	
	\begin{eqnarray}  \label{mast}
		\dot{\rho} &=&i[\rho, H]+\frac{\kappa }{2}\sum_{j=\circlearrowleft, 
			\circlearrowright}\mathcal{L}[a_{j}]\rho +\frac{\gamma 
			_{m}}{2}\bar{n}_{m} \mathcal{L}[b^{\dag }]\rho \nonumber\\
		&&+\frac{\gamma _{m}}{2}(
		\bar{n}_{m}+1)\mathcal{L}[b]\rho, 
	\end{eqnarray}
	where $\mathcal{L}[o]\rho \equiv 2o\rho o^{\dagger } - (o^{\dagger }o\rho + \rho o^{\dagger }o)$ defines the Lindblad dissipator for operator $o \in \{a_j, a^{\dagger}_j, b, b^{\dagger}\}$, and $\rho$ represents the density matrix of the complete system. The parameters $\kappa$ and $\gamma_m$ correspond to the decay rates of the optical cavity field and mechanical resonator, respectively. The thermal occupation of the mechanical mode is given by $\bar{n}_m = [\exp(\omega_m/k_B T) - 1]^{-1}$, where $k_B$ is the Boltzmann constant and $T$ denotes the environmental temperature.

	To more effectively mitigate the quantum effects induced by radiation pressure, we perform displacement operations on both the density operator and the field operator, defined as $D_j(\alpha_j)=\exp(\alpha_j a_j^{\dagger}-\alpha_j^{*}a_j)$ and $D_b(\beta)=\exp(\beta b^{\dagger}-\beta^{*}b)$. Here, $\alpha_j$ and $\beta$ represent the displacement parameters for the optical mode and the mechanical oscillator mode, respectively. This transformation simplifies the analysis by shifting the operators to a new reference frame, facilitating a more tractable description of the system dynamics~\cite{PhysRevA.109.063508, PhysRevA.110.023519}. 
	Within the linearization framework, we derive the master equation in the displacement representation as follows:
	\begin{eqnarray}
		\dot{\rho}' &=&i[\rho', H_{\rm eff}]+\frac{\kappa }{2}\sum
		_{j=\circlearrowleft, 
			\circlearrowright}\mathcal{L}[a_{j}]\rho' +\frac{\gamma 
			_{m}}{2}\bar{n}_{m} \mathcal{L}[b^{\dag }]\rho' \nonumber\\
		&&+\frac{\gamma 
			_{m}}{2}(%
		\bar{n}_{m}+1)\mathcal{L}[b]\rho'
	\end{eqnarray}
   with $\rho'(t)= \Pi_{j}D_j(\alpha_j)D_b(\beta)\rho(t) 
	D^{\dagger}_b(\beta)\Pi_j D^{\dagger}_j(\alpha_j)$ for $j=\circlearrowleft, 
	\circlearrowright$.
	Here, the system dynamics is governed by the effective Hamiltonian:
	\begin{eqnarray}
		H_{\text{eff}} &=& \omega_m b^{\dagger}b + \sum_{j= \circlearrowleft, 
			\circlearrowright}\tilde{\Delta}_j 
		a_j^{\dagger}a_j - (G_j^{\ast}a_j + G_ja_j^{\dagger})(b^{\dagger}+b) 
		\nonumber\\
		&& + J(a_{\circlearrowright}^{\dagger}a_{\circlearrowleft}
		+ a_{\circlearrowleft}^{\dagger}a_{\circlearrowright}).
	\end{eqnarray} 
	The normalized driving detuning $\tilde{\Delta}_j$ is defined as 
	$\tilde{\Delta}_j 
	= \Delta_j - 2g_0\Re[\beta]$ with $\Delta_j = \omega_j - \omega_L$, and 
	$G_j = g_0 \alpha_j$.
	The classical motion amplitudes $\alpha_j(t)$ and $\beta(t)$ are governed 
	by 	the equations of motion:
	\begin{subequations}\label{alpha}
		\begin{align}
			\dot{\alpha}_{\circlearrowleft}(t) &= 
			-(i\tilde{\Delta}_{\circlearrowleft}  + 
			\kappa/2)\alpha_{\circlearrowleft} - 
			iJ\alpha_{\circlearrowright} + \varepsilon, \\
			\dot{\alpha}_{\circlearrowright}(t) &= 
			-(i\tilde{\Delta}_{\circlearrowright} + 
			\kappa/2)\alpha_{\circlearrowright} 
			- iJ\alpha_{\circlearrowleft}, \\
			\dot{\beta}(t) &= -(i\omega_m + \gamma_m/2)\beta + 
			ig_0(|\alpha_{\circlearrowleft}|^2 + 
			|\alpha_{\circlearrowright}|^2).
		\end{align}
	\end{subequations}
	The dynamical evolution of  the
	second-order moments can be studied using the covariance method based on 
	quantum master equation
	$\langle \dot{O}\rangle ={\rm Tr}[O\dot{\rho}] $
	
	\begin{equation} \label{xx}
		\dot{\mathbf{X}}= {\mathbf{A}}(t)\cdot {\mathbf X} + {\mathbf D}(t),
	\end{equation}
	where  ${\bf{X}}=[x_0, x_1, ..., x_{20}]$
	and $\mathbf{A}(t)$ is the time-dependent matrix (see Appendix~\ref{appa} for 
	details). 
	The inhomogeneous term $\mathbf{D}(t)$ depends on 
	$G_{\circlearrowleft}(t)$ and $G_{\circlearrowright}(t)$, and is expressed 
	as:
	\begin{eqnarray}
		&&\mathbf{D}(t)=[0, 0, 
		\gamma_m\bar{n}_m, 0,0, 
		0, 0, 
		0,0,0,0, 0,0, 0,0,\nonumber\\
		&& -iG_{\circlearrowleft}^{*}(t), 
		iG_{\circlearrowleft}(t), 0, 0, -iG_{\circlearrowright}^{*}(t), 
		iG_{\circlearrowright}(t)].
	\end{eqnarray}

	\section{Quantum Fisher information of two-mode Gaussian state}\label{III}
	
   We estimate the rotation angular velocity \(\Omega\) of the COM resonator by analyzing the output optical field and employ quantum parameter estimation theory to quantify the precision.
   According to the quantum Cramér-Rao bound~\cite{Helstrom1969, Rao1992}, the minimum uncertainty in estimating \(\Omega\) is given by \(
    \Delta \Omega_{\rm min} ={1}/{\sqrt{\nu \mathcal{F}_{\Omega}}},\)
    where \(\mathcal{F}_{\Omega}\) represents the quantum Fisher information (QFI) associated with \(\Omega\), and \(\nu\) denotes the number of repeated measurements (hereafter, we set \(\nu = 1\)). 
    The QFI quantifies the system's sensitivity to changes in \(\Omega\), with higher values indicating better precision. In quantum metrology, precision is bounded by the SQL, \(\mathcal{F} \sim N\), and the Heisenberg limit (HL), \(\mathcal{F} \sim N^2\), where \(N\) is the total number of particles or resources. When \(\mathcal{F} / N \gg 1\), precision surpasses the SQL, showcasing the advantage of quantum-enhanced metrology for superior sensitivity.

	For Gaussian states the QFI  depends on the displacement vector and the covariant matrix. The elements of the displacement vector $\bf d$ and the covariant matrix 
	$\bf \sigma$ 
	are defined as ${\bf d}_i={\rm Tr}[\rho(t)\hat{ A}_i]$, and 
	${\bf \sigma}_{ij}={\rm Tr}[\rho(t)\{\Delta 
	\hat{A}_i, \Delta \hat{A}^{\dagger}_j\}]$, with $\hat{A}=(\hat{a}_{ 
		\circlearrowleft}, \hat{a}_{ \circlearrowleft}, 
	\hat{a}^{\dagger}_{{ \circlearrowleft}}, \hat{a}^{\dagger}_{{ 
			\circlearrowright}} )^{T}$ and $\Delta \hat{A}_i 
	=\hat{A}_i -d_i$. The QFI with respect to the measured quantity $\Omega$ in 
	the bimodal Gaussian state $\rho(t)$ can be calculated via~\cite{	
		Safranek_2019,Liu2020}
	\begin{equation} \label{qfi}
		\mathcal{F}_{\Omega} = \frac{1}{2}[{\rm vec}(\partial_{\Omega} 
		\sigma)]^{\dagger} 
		{\mathbf 
			M^{-1} }
		{\rm vec}(\partial_{\Omega} \sigma) +2(\partial_{\Omega} {\bf 
			d})^{\dagger}\sigma^{-1}(\partial_{\Omega} {\bf d}),
	\end{equation}
	where ${\mathbf M}= \sigma^{*}\otimes \sigma -K\otimes K$ with $\sigma^{*}$ 
	being the  
	complex conjugate of $\sigma$ and 
	$K = {\rm diag}(1,1, -1, -1)$, 
	and $\rm vec[\rm m]$ denotes vectorization of a matrix, which is 
	defined as a column vector constructed from columns of a matrix as
	${\rm m}=\begin{bmatrix}
		a & b \\
		c & d \\
	\end{bmatrix}$,   ${\rm vec}[{\rm m}]=(a, c, b,d)^{T}$.
	However, the expression of Eq.~(\ref{qfi}) cannot be used when at least one of the 
	modes is in a pure state (at least not without any modification), because 
	${\mathbf M}$ is not noninvertible in that case, then a convenient way 
	to calculate the QFI of pure states is
	\begin{equation}\label{qfi2}
		\mathcal{F}_{\Omega} = \frac{1}{4} {\rm 
			Tr}(\sigma^{-1}\partial_{\Omega} \sigma
		\sigma^{-1}\partial_{\Omega} \sigma)
		+2(\partial_{\Omega} {\bf d})^{\dagger}\sigma^{-1}\partial_{\Omega}
		{\bf d}.
	\end{equation}
	
    For our analysis of Gaussian states, we first return to the state \(\rho\) prior to the displacement operation. By applying the inverse displacement operation with respect to \(\rho'\), we derive the displacement vectors
    \( {\bf d} = \left[\alpha_{\circlearrowleft}, \alpha_{\circlearrowright}, \alpha^{*}_{\circlearrowleft}, \alpha^{*}_{\circlearrowright}\right] \),
    which are determined by solving Eq.~(\ref{alpha}). These vectors depend on the parameter
    \(\Omega\) and  their derivatives \(\partial_{\Omega} {\bf d} \)  can be obtained through numerical calculations.
    For Gaussian states, the corresponding covariance matrix \(\sigma\) remains unchanged under the displacement operation. Utilizing Eq.~(\ref{qfi}), the calculation of the QFI additionally requires knowledge of both the covariance matrix \(\bf \sigma\) and its derivative with respect to the parameter to be measured, denoted as \(\partial_{\Omega} \sigma\).
	The dynamics of 
	$\partial_{\Omega} \sigma$, governed by the following equation
	\begin{equation} \label{aequ}
		\begin{bmatrix}
			\dot{\bf X} \\
			\partial_{\Omega}\dot{\bf X}
		\end{bmatrix}
		=
		\begin{bmatrix}
			{\bf A}(t) & {\bf 0}_{21\times 21} \\
			\partial_{\Omega}{\bf A}(t) & {\bf A}(t)
		\end{bmatrix}
		\begin{bmatrix}
			{\bf X} \\
			\partial_{\Omega}{\bf X}
		\end{bmatrix} 
		+\begin{bmatrix}
			{\bf D} \\
			\partial_{\Omega}{\bf D}
		\end{bmatrix}.
	\end{equation}

\begin{figure*}[tp]
	\centering\includegraphics[width=17.8cm]{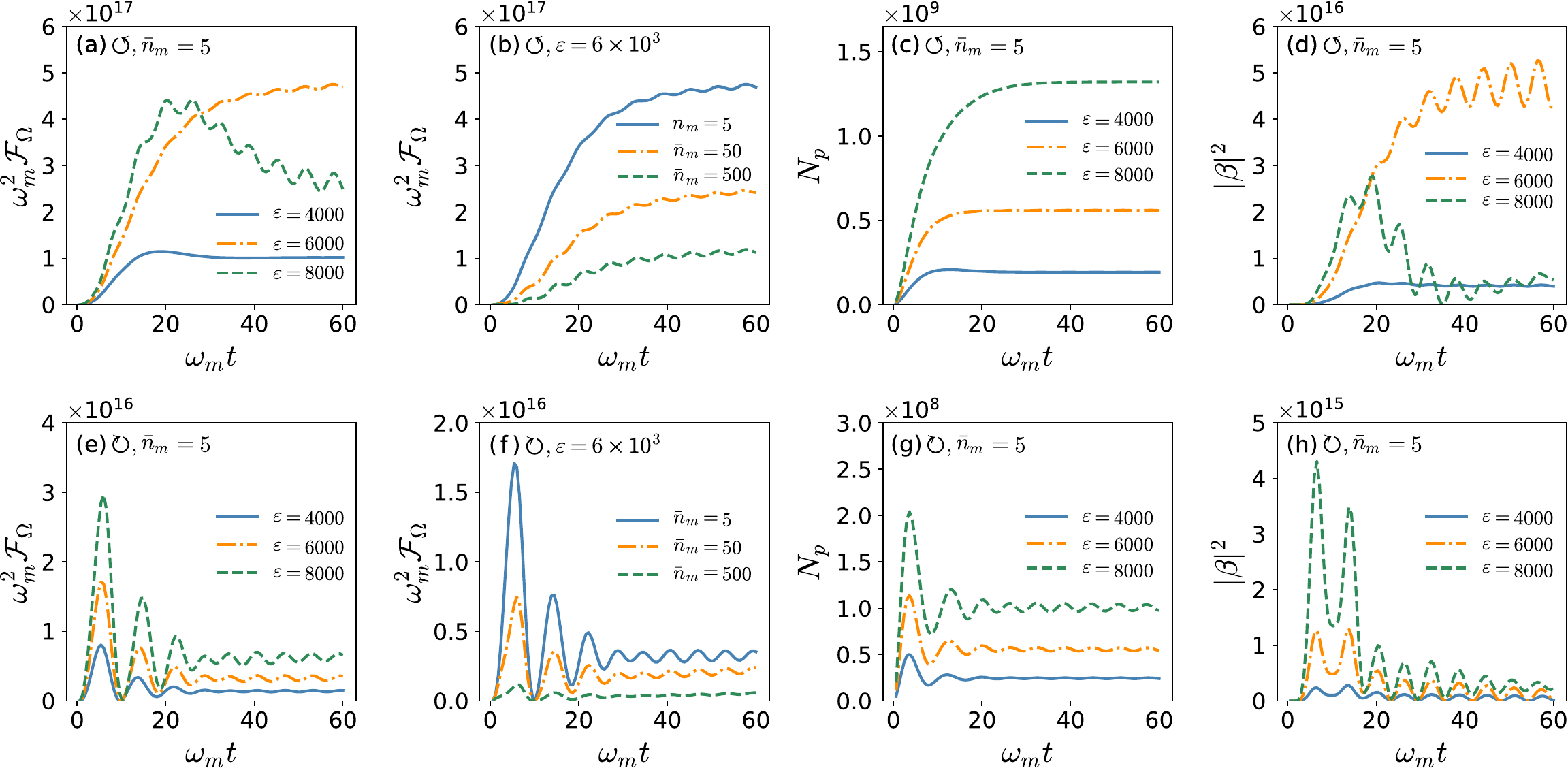}
	\caption{
		(a) and (b) Dynamics of the QFI \(\mathcal{F}_{\Omega}\) in the CCW configuration for different values of \(\varepsilon\) and thermal phonon number \(\bar{n}_{m}\) in the vibrational mode.
		(c) and (d)) Dynamics of the total intracavity photon number \(N_p = |\alpha_{\circlearrowleft}|^2 + |\alpha_{\circlearrowright}|^2\) and the total phonon number \(|\beta|^2\) in the CCW configuration for different values of \(\varepsilon\).
		(e) and (f) Dynamics of the QFI \(\mathcal{F}_{\Omega}\) in the CW configuration for different values of \(\varepsilon\) and thermal phonon number \(\bar{n}_{m}\) in the vibrational mode.
		(g) and (h) Dynamics of the total intracavity photon number \(N_p = |\alpha_{\circlearrowleft}|^2 + |\alpha_{\circlearrowright}|^2\) and the total phonon number \(|\beta|^2\) in the CW configuration for different values of \(\varepsilon\).
		Parameters are as follows: \(\kappa = 0.44\omega_m\), \(\gamma_m = 3.5 \times 10^{-3}\omega_m\), \(\Omega = 2\text{ kHz}\), \(\Delta_c/\omega_m = 0.5\), and \(J = 0\).
	} \label{Fig2}
\end{figure*}
	
	By numerically solving the above equations and incorporating 
	the input-output relation $a_{i,  \rm out}(t) =\sqrt{\kappa}  a_i(t) 
	-a_{i,\rm in}(t)$ 
	($i=\circlearrowleft, \circlearrowright$; see Appendix~\ref{appb}),
	we can obtain the matrix $\sigma_{\text{out}}$ and 
	$\partial_{\Omega}\sigma_{\text{out}}$
	which have the following structure:
	\begin{subequations}
		\begin{align}
			{\bf \sigma}_{\rm out}(t)&=\begin{bmatrix}
				X_{\rm out}(t) & Y_{\rm out}(t) \\
				Y^{*}_{\rm out}(t) & X^{*}_{\rm out}(t) \\
			\end{bmatrix}, \\
			\partial_{\Omega} {\bf \sigma}_{\rm out}(t)&=\begin{bmatrix}
				\partial_{\Omega}X_{\rm out}(t) &  \partial_{\Omega}Y_{\rm 
					out}(t) \\
				\partial_{\Omega}Y^{*}_{\rm out}(t) & 
				\partial_{\Omega}X^{*}_{\rm out}(t) \\
			\end{bmatrix}	
		\end{align}	
	\end{subequations}
	with
	\begin{subequations}\label{aout}
		\begin{align}
			X_{\rm out}&=2\begin{bmatrix}
				\langle a_{\circlearrowleft,{\rm out} }^{\dag 
				}a_{\circlearrowleft,{\rm out}
				}\rangle +1/2 & \langle a_{\circlearrowright,{\rm out} }^{\dag
				}a_{\circlearrowleft,{\rm out} }\rangle \\ 
				\langle a_{\circlearrowleft, {\rm out} }^{\dag 
				}a_{\circlearrowright,{\rm out}
				}\rangle & \langle a_{\circlearrowright, {\rm out} }^{\dag
				}a_{\circlearrowright,{\rm out} }\rangle +1/2
			\end{bmatrix}, \\
			Y_{\rm out} &=
			2\begin{bmatrix}
				\left\langle a_{\circlearrowleft,{\rm out} 
				}a_{\circlearrowleft,{\rm out} }\right\rangle
				& \left\langle a_{\circlearrowleft,{\rm out} 
				}a_{\circlearrowright, {\rm out} }\right\rangle
				\\ 
				\left\langle a_{\circlearrowleft, {\rm out} 
				}a_{\circlearrowright,{\rm out} }\right\rangle
				& \left\langle a_{\circlearrowright, {\rm out} 
				}a_{\circlearrowright, {\rm out} }\right\rangle
			\end{bmatrix},
		\end{align}	
	\end{subequations}

	In Eq.~(\ref{aout}), the expectation values of the output field can be experimentally obtained via homodyne detection~\cite{Riedinger2018,Ockeloen-Korppi2018,doi:10.1126/science.abf2998,PhysRevA.95.042322} (see Appendix~\ref{appc} for details),  and are mathematically related to the input field as given by:
	\begin{subequations} \label{ain}
		\begin{align}
			\langle a^{\dagger}_{i,  \rm out}(t) a_{j,  \rm out}(t) \rangle 
			&={\kappa}  \langle a^{\dagger}_{i}(t) a_{j}(t) \rangle - 
			\sqrt{\kappa}\langle a^{\dagger}_{i}(t) a_{j,  \rm in}(t) \rangle  
			\nonumber\\
			&- \sqrt{\kappa}\langle a^{\dagger}_{i,  \rm in}(t) a_{j}(t)\rangle 
			+ 
			\langle 
			a^{\dagger}_{i,  \rm in}(t) a_{j,  \rm in}(t) \rangle,  \\
			\langle a_{i, \rm out}(t) a_{j,  \rm out}(t) \rangle 
			&={\kappa}  \langle a_{i}(t) a_{j}(t) \rangle  - 
			\sqrt{\kappa}\langle 
			a_{i}(t) a_{j,  \rm in}(t) \rangle \nonumber \\
			&-\sqrt{\kappa}\langle a_{i,  \rm in}(t) a_{j}(t) \rangle + \langle 
			a_{i,\rm in}(t) a_{j,  \rm in}(t) \rangle. \nonumber\\
		\end{align}
	\end{subequations}
	For optical modes coupled to vacuum fields, as described in Eq.~(\ref{mast}), the expectation values satisfy  
	$\langle a^{\dagger}_{i, \rm in}(t) a_{j, \rm in}(t) \rangle =  
	\langle a_{i, \rm in}(t) a_{j, \rm in}(t) \rangle = 0. $
	The terms of 
	$\langle a^{\dagger}_{i}(t) a_{j}(t) \rangle$ and  $\langle a_{i}(t) 
	a_{j}(t) \rangle$ can be obtained by 
	numerically solving Eq.~(\ref{aequ}). However, in Eq.~(\ref{ain}) there are also
	contain expectation values involving products of input and system 
	operators, such as  $\langle a^{\dagger}_{i}(t) a_{j,  \rm in}(t) \rangle$, 
	$\langle a^{\dagger}_{i,  \rm in}(t) a_{j}(t) \rangle$, $\langle a_{i}(t) 
	a_{j,  \rm in}(t) \rangle$
	and $\langle a_{i, \rm in}(t) a_{j}(t) \rangle$, these terms
	are in general different from zero and cannot be directly calculated by the 
	master equation approach, which does not calculate mixed bath-system 
	correlations.  This problem can be solved by deriving the commutation 
	relations between
	system and input operators $[a(t), a_{\rm in}^{(\dagger)}(s)]= 
	\sqrt{\kappa}u(t-s)[a(t),a^{(\dagger)}(s)]$ where $u(t-s)$ is equal to 1 if 
	$t > s$, $1/2$ if $t = s$, and 0 if $t < s$. Following  the calculation 
	procedure of Ref.~\cite{PhysRevA.30.1386} and choosing $t=s$, we therefore have 
	$\langle a_{i}(t) a_{j,  \rm in}^{\dagger}(t) \rangle = 
	\frac{\sqrt{\kappa}}{2} \delta_{ij}$, $\langle  a_{i,  \rm 
		in}^{\dagger}(t)a_{j}(t) \rangle = 
	0$ and $\langle  a_{i,  \rm 
		in}(t)a_{j}(t) \rangle = 0$.
	
	With the help of Eqs.~(\ref{qfi}-\ref{ain}), we can then obtain the specific value of QFI 
	$\mathcal{F}_{\Omega}$.

	\section{ NonReciprocal Optomechanical Gyroscope} \label{IV}
	
	As studied in Ref.~\cite{PhysRevLett.125.143605},  the rotation of optical microcavities 
	generates the Sagnac effect, leading to frequency shifts in opposite 
	directions that break the time-reversal symmetry of the system, resulting 
	in quantum non-reciprocity. 
	Here, we delve into the non-reciprocal properties of the optomechanical 
	gyroscope and their influence on the estimation accuracy of the COM 
	resonator's  rotation angular velocity $\Omega$. Utilizing realistic parameters derived 
	from recent experiments~\cite{Maayani2018, Righini2011}, specifically $n=1.48$, 
	$m=10$ng, $R=1.1$mm, 
	$\lambda=0.78\mu$m, $\omega_m=50$MHz,  $\kappa=22$MHz and $\gamma_m = 175$kHz, we aim to offer a comprehensive 
	understanding of the gyroscope's behavioral characteristics. Moreover, we 
	explore strategies aimed at effectively enhancing the sensitivity of 
	$\Omega$ estimation.

	In Fig.~\ref{Fig2}, we employ numerical methods to investigate the non-reciprocal dynamic behavior of the QFI \(\mathcal{F}_{\Omega}\), which is derived from the displacement vector and the second-order correlation function of the output field by solving Eqs.~(\ref{qfi}-\ref{ain}).  Here, we first define the rotation angular velocity $\Omega$ of the COM resonator as positive for CW rotation. We consider both CCW and CW driving fields in the COM system, corresponding to left-side  (as illustrated in Fig.~\ref{Fig1}) and right-side driving, respectively.
	We maintain a fixed pump-cavity detuning (\(\Delta_c = \omega_c - \omega_L\)) of \(\Delta_c/\omega_m = 0.5\) and an angular frequency of \(\Omega = 2\) kHz.
    By comparing Figs.~\ref{Fig2}(a–b) and \ref{Fig2}(e–f), we observe that the dynamical behavior of the QFI exhibits pronounced non-reciprocity. Specifically, in the CCW scenario, the QFI is more than one order of magnitude higher than in the CW case. Moreover, the CCW and CW dynamics show distinct characteristics. In the CCW case, when \(\varepsilon\) is relatively small (\(\varepsilon < 6000\)), \(\mathcal{F}_{\Omega}\) initially increases monotonically before stabilizing at its maximum. For sufficiently large \(\varepsilon\), \(\mathcal{F}_{\Omega}\) instead exhibits oscillatory behavior with a subsequent decreasing trend. In contrast, in the CW scenario, \(\mathcal{F}_{\Omega}\) demonstrates oscillatory dynamics for all considered values of \(\varepsilon\), eventually stabilizing at a relatively low value regardless of \(\varepsilon\).

    Furthermore, a comparison of Figs.~\ref{Fig2}(b) and \ref{Fig2}(f) reveals that increasing the thermal phonon number \(\bar{n}_m\) reduces the value of \(\mathcal{F}_{\Omega}\), indicating that strong dissipation suppresses the QFI for both the CCW and CW configurations. Additionally, Figs.~\ref{Fig2}(c) and \ref{Fig2}(g) illustrate the dynamics of the total intracavity photon number \(N_p = |\alpha_{\circlearrowleft}|^2 + |\alpha_{\circlearrowright}|^2\) for the CCW and CW cases, respectively. These results demonstrate that larger values of \(\varepsilon\) lead to higher excitation of the total photon number, which eventually stabilizes. Notably, the relationship \(\mathcal{F}_{\Omega} \sim N_p^2\) emerges due to quantum correlations between the two distinct optical modes~\cite{PhysRevLett.125.143605}, indicating HL scaling when the total probe field is treated as the resource.

    However, in Figs.~\ref{Fig2}(d) and \ref{Fig2}(h), we observe that for the phonon number \(|\beta|^2\) of the vibrational mode, large values of \(\varepsilon\) (e.g., \(\varepsilon > 6000\)) induce significant oscillations, exhibiting a trend similar to that of the QFI. This suggests that although \(|\beta|^2\) is not a direct observable in the QFI calculation, it still influences the QFI by modulating the covariance matrix of the optical field. Even when considering both \(N_p\) and \(|\beta|^2\) as the total resources consumed to enhance \(\Omega\), the condition \(\mathcal{F}_{\Omega}/(N_p + |\beta|^2) \gg 1\) holds, demonstrating that, in this sense, our gyroscope scheme still surpasses the SQL.
    
    From Fig.~\ref{Fig2}, we further observe that the CCW configuration achieves a higher QFI value by exciting a larger number of photons and phonons compared to the CW case  for the considered detuning parameter  $\Delta_c$. These findings underscore the critical role of optimizing the pump field direction for fixed detuning $\Delta_c >0$, increasing its intensity, and tuning the thermal phonon population of the vibrational mode to significantly enhance the sensitivity of non-reciprocal gyroscopic measurements.
    
\begin{figure*}[tp]
    	\centering\includegraphics[width=16cm]{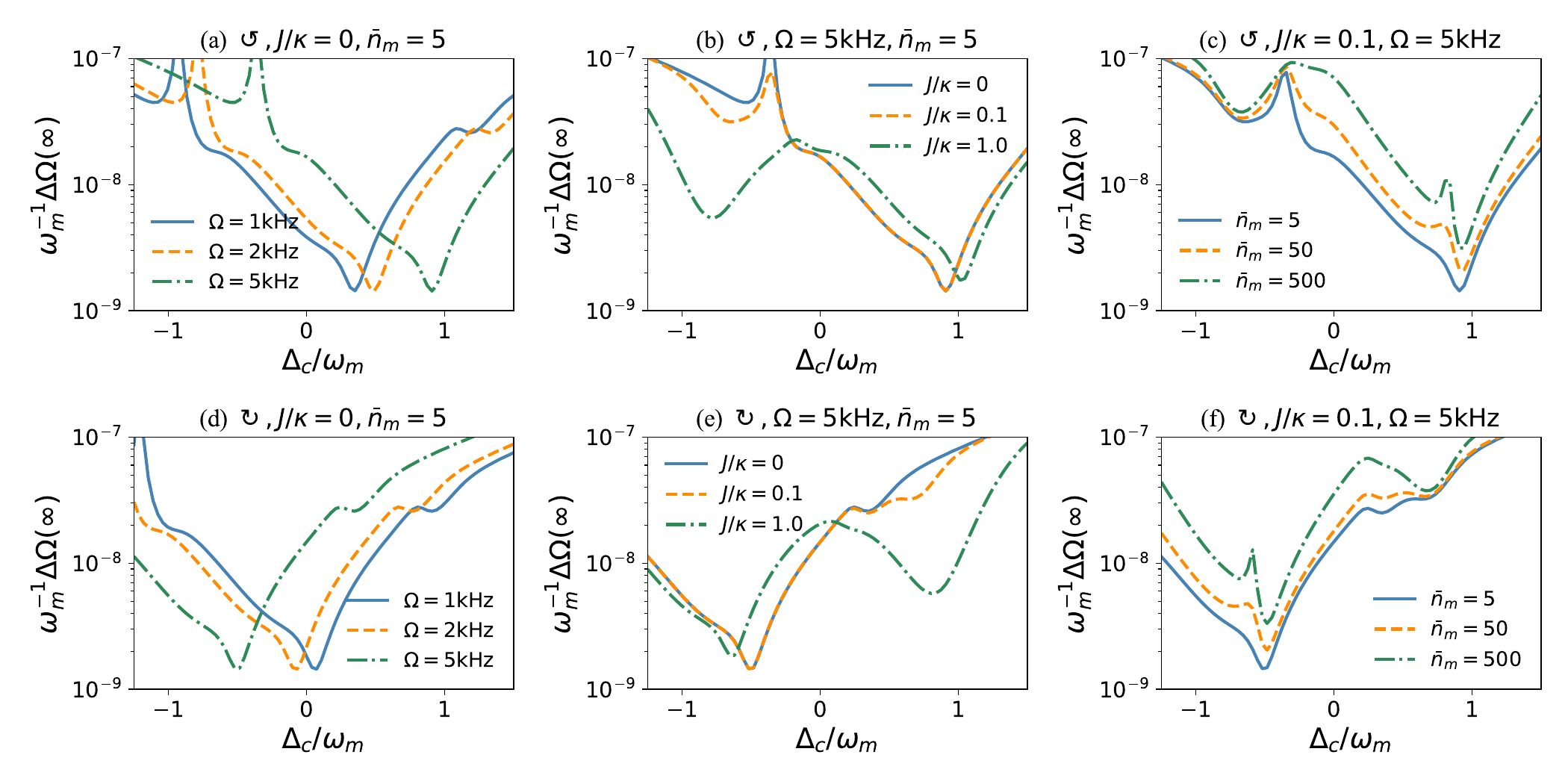}
    	\caption{ 
    		(a)-(c) Variation of steady-state estimation error $\Delta{\Omega}(\infty)$ with pump-cavity detuning $\Delta_c/\omega_m$ in the CCW configuration.
    		The steady-state estimation error $\Delta{\Omega}(\infty)$ is plotted as a function of the pump-cavity detuning $\Delta_c/\omega_m$ in the CCW configuration, for different values of the angular velocity $\Omega$, coupling strength $J$, and thermal phonon number $\bar{n}_m$ in the environment.
    		(d)-(f) Variation of steady-state estimation error $\Delta{\Omega}(\infty)$ with pump-cavity detuning $\Delta_c/\omega_m$ in the CW configuration.
    		The steady-state estimation error $\Delta{\Omega}(\infty)$ is similarly plotted for the CW configuration, under the same variations of $\Omega$, $J$, and $\bar{n}_m$.
    		Parameters are  as follows: $\kappa = 0.44\omega_m, \gamma_m = 3.5\times 10^{-3}\omega_m$ and $\varepsilon=6\times10^3$.
    	} \label{Fig3}
    \end{figure*}

    To systematically investigate the influence of pump-cavity detuning on angular velocity sensing precision and its role in establishing nonreciprocity, Fig.~\ref{Fig3} demonstrates the steady-state measurement precision \(\Delta \Omega(\infty)\) as a function of normalized detuning \(\Delta_c/\omega_m\) for different angular velocities \(\Omega\). This analysis is performed at a fixed driving strength \(\varepsilon = 6000\), which guarantees the QFI has reached its saturation regime.
    From the figure, it is evident that the optimal value of \(\omega^{-1}_m\Delta \Omega(\infty) = 1/\sqrt{\omega^2_m \mathcal{F}_{\Omega}}\) can approach \(10^{-9}\), indicating that the absolute angular velocity measurement precision \(\Delta \Omega_{\rm min}\) can reach \(10^{-3}\) Hz, with a relative error \(\Delta \Omega_{\rm min}/\Omega \sim 10^{-6}\).
    
    Figure~\ref{Fig3} also demonstrates that the highest precision \(\Delta \Omega(\infty)\) is achieved only at specific values of \(\Delta_c/\omega_m\), which vary with \(\Omega\). In the CCW scenario, the optimal precision occurs in the \(\Delta_c/\omega_m > 0\) region, whereas in the CW case, the peak appears in the opposite region. Notably, these peaks exhibit symmetric distributions around \(\Delta_c/\omega_m \approx 0.1\).
    
    Furthermore, Figs.~\ref{Fig3}(b, e) and \ref{Fig3}(c, f) reveal that both the backscattering strength \(J\) and thermal phonon number \(\bar{n}_m\) significantly suppress the peak precision. Notably, a higher \(J\) value (e.g., \(J/\kappa=1\)) not only reduces the optimal precision but also induces a double-peak structure, underscoring its crucial role in governing gyroscopic sensitivity.
    
    A comprehensive analysis of Figs.~\ref{Fig2} and \ref{Fig3} reveals pronounced non-reciprocal behavior in the gyroscope, indicating that its sensitivity can be significantly enhanced by selecting an appropriate direction for  laser pump field. Moreover, the sensitivity is strongly dependent on the detuning parameter \(\Delta_c\), with the asymptotic behavior of \(\Delta\Omega(\infty)\) exhibiting distinct local characteristics. Notably, for a fixed \(\Delta_c\), both the optimal QFI and the minimum achievable uncertainty \(\Delta\Omega\) are highly sensitive to the angular velocity \(\Omega\), presenting a critical challenge: achieving improved measurement precision across a wide range of unknown angular velocities remains unresolved.

    To overcome this challenge, we introduce in the following section a  RL-based framework for dynamic gyroscopic sensitivity optimization, preserving the CCW driving field configuration through left-side optical pumping. This methodology ensures uniform sensitivity optimization for both positive (\(\Omega>0\), CW) and negative (\(\Omega<0\), CCW) angular velocities.  Owing to symmetry, the case of $\Omega < 0$ is mathematically equivalent to that of $\Omega > 0$ when the driving field is applied through the right-side input.  This equivalence significantly suppresses nonreciprocal effects that would otherwise degrade the accuracy of angular velocity measurements.
    
    Through intelligent agent training for adaptive adjustment of \(\Delta_c\), our approach achieves robust measurement consistency across varying operational conditions. This advancement not only extends the performance limits of optomechanical gyroscopes but also provides fundamental insights applicable to precision measurement systems and quantum sensing technologies.

	\section{Reinforcement learning assisted global optomechanical gyroscope} \label{V}
	
     To enhance the versatility and performance of the optomechanical gyroscope across a broad range of angular velocities, it is essential to develop a strategy that maximizes the average QFI, denoted as \(\bar{\mathcal{F}}_{\Omega}\), within a target angular velocity range. Specifically, we define \(\bar{\mathcal{F}}_{\Omega}\) over the interval \(\Omega \in [\Omega_{\rm min}, \Omega_{\rm max}]\) as:
     \begin{eqnarray}\label{aqfi}
     \bar{\mathcal{F}}_{\Omega} = \frac{1}{\Omega_{\rm max} - \Omega_{\rm min}} \int_{\Omega_{\rm min}}^{\Omega_{\rm max}} \mathcal{F}_{\Omega} \, d\Omega.
     \end{eqnarray}
     To achieve this goal, we employ RL approach to optimize the detuning parameter \(\Delta_c = \omega_c - \omega_L\), which characterizes the frequency difference between the cavity mode (\(\omega_c\)) and the laser mode (\(\omega_L\)). Our primary objective is to maximize \(\bar{\mathcal{F}}_{\Omega}\), thereby ensuring optimal sensitivity and precision of the gyroscope across the specified angular velocity range.

	\begin{figure*}[tp]
		\centering\includegraphics[width=15cm]{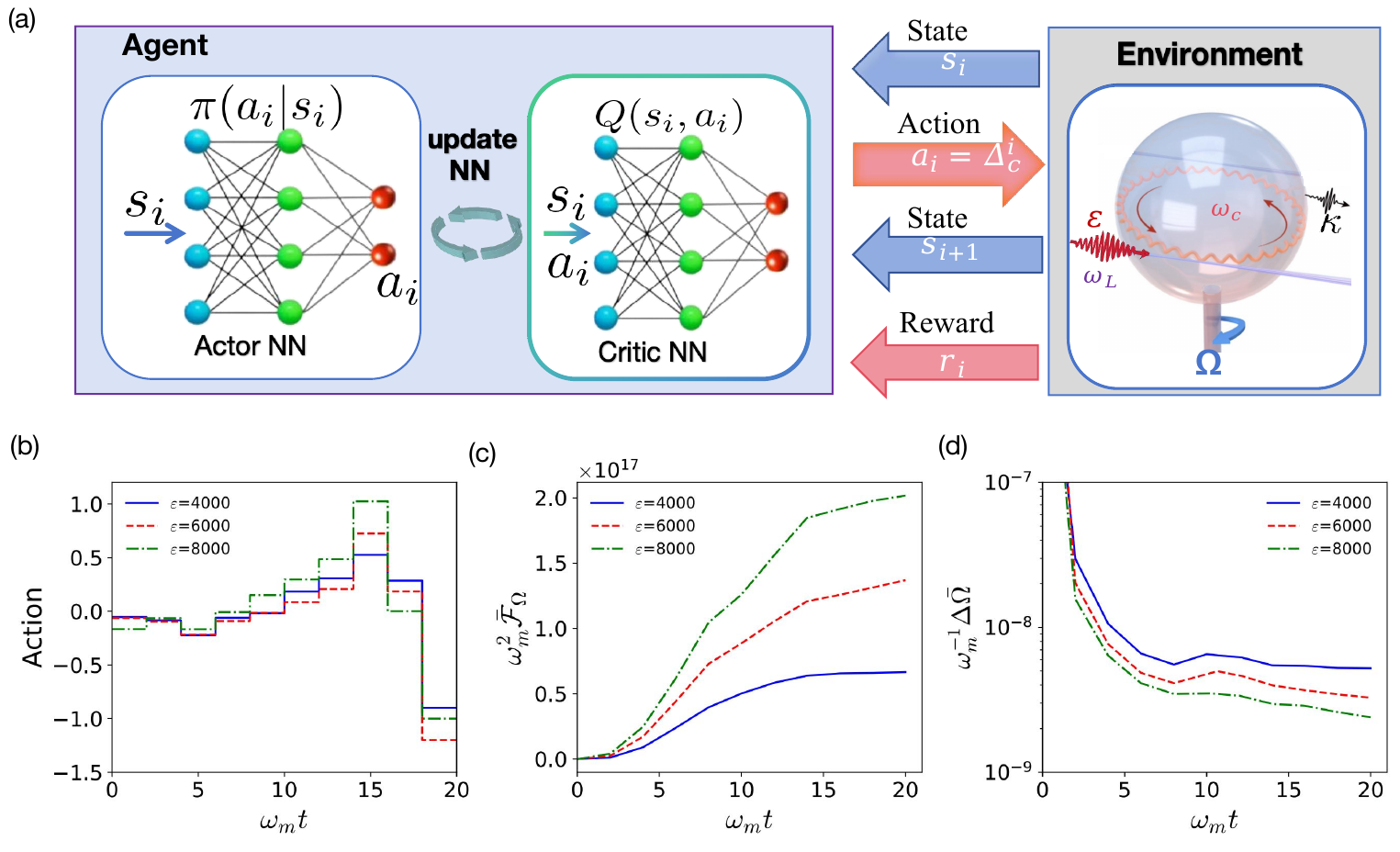}
		\caption{ (a) Schematic representation of the PPO learning process.
			The PPO agent comprises two neural networks (NNs): the actor NN and the critic NN. At each time step \(i\), the actor NN selects an action \(a_i = (\Delta^i_c)\) based on the current environment (optomechanical gyroscope) state \(s_i\) through the policy \(\pi(a_i|s_i)\). The environment, modeled as COM system, evolves to a new state \(s_{i+1}\) and provides a reward \(r_i\) to the agent. The agent uses this information (state, action, reward, and new state) to update both the actor and critic NNs, optimizing the policy \(\pi(a|s)\) and value functions \(Q(s, a)\) for improved decision-making.
			(b)-(d) Optimal performance of the global optomechanical gyroscope across the frequency range \(\Omega \in [-4, 4]\)kHz when operating with a CCW driving field configuration (left-side pump).
			(b) The optimal control sequences \(\Delta_c/\omega_m\) for maximizing the reward function were derived through the PPO algorithm, with distinct parameter configurations corresponding to different values of \(\varepsilon\).
			(c) Dynamics of the optimal average QFI \(\mathcal{\bar{F}}_{\Omega}\) under the optimal action sequence, depicted as a function of the parameter \(\varepsilon\).
			(d) Optimal average estimation error \(\Delta \bar{\Omega}\) under the optimal action sequence, presented as a function of \(\varepsilon\).
			Parameters are as follows: \(\kappa = 0.44\omega_m\), \(\gamma_m = 3.5 \times 10^{-3}\omega_m\), \(\bar{n}_{m} = 5\), and \(J/\kappa = 0.1\).
            } \label{Fig4}
		
	\end{figure*}

	In a conventional RL framework, the interaction between the agent and the  environment is governed by three key components: the state space \(\mathcal{S}\), the action space \(\mathcal{A}\), and the reward space \(\mathcal{R}\). This interaction is modeled as a finite Markov decision process (MDP), where the total training time \(\tau\) is divided into \(n\) discrete steps, each with a fixed interval \(\delta \tau = \tau/n\). At each time step \(t\), the agent observes the current state \(s_t \in \mathcal{S}\) and selects an action \(a_t\) based on the policy \(\pi(\mathcal{A}|\mathcal{S})\). Executing the action transitions the system to a new state \(s_{t+1} \in \mathcal{S}\), and the agent receives a reward \(r_{t+1}\) from the reward space \(\mathcal{R}\). This Markovian process is described as a sequence of events: \((s_0, a_0, r_1, s_1, \ldots, s_t, a_t, r_{t+1}, s_{t+1}, \ldots, s_{n-1}, a_{n-1}, r_n, s_n)\), where \(n\) represents the total number of steps in an episode~\cite{PhysRevLett.131.050601, Xiao2022,PhysRevA.109.042417}.
	
	For policy optimization, we adopt the Proximal Policy Optimization (PPO) algorithm~\cite{PPO}, a state-of-the-art actor-critic method that has proven to be highly effective for maximizing cumulative rewards \(\mathcal{R}\). PPO is particularly suited for our application due to its robustness, sample efficiency, and ability to manage continuous action spaces, making it ideal for optimizing complex systems like the optomechanical gyroscope. The first and second moments of the total system's density operator \(\rho_{\rm tot}\) serve as the input state for the agent. These moments encapsulate comprehensive information about the quantum state, providing crucial data for guiding the agent’s decision-making process.
	
	The PPO algorithm is comprised of two key components: the actor neural network (NN) and the critic NN, which collaboratively work to optimize the agent's policy. The actor NN receives the environment state \(s_i\) as input and outputs an action \(a_i\) through the policy function \(\pi(a_i|s_i)\). In this system, \(a_i = \Delta_c^i\) represents the pump-cavity detuning, which is applied to the environment. The actor NN continuously updates its policy to maximize the long-term cumulative reward. Simultaneously, the critic NN evaluates the value of taking action \(a_i\) in state \(S_i\) using the value function \(Q(s_i, a_i)\), providing feedback that helps refine the policy of the actor NN, leading to higher rewards. This iterative process of evaluation and optimization ensures that the agent's decision-making strategy improves over time.
	
	In our framework, the environment represents the physical system of the optomechanical gyroscope, where the agent’s action \(a_i\) influences the system’s state, transitioning from \(s_i\) to \(s_{i+1}\). Based on the agent’s action, the environment generates a reward \(r_i\) that guides further policy updates. This feedback loop enables the agent to adapt its strategy, maximizing the reward and improving system performance.
	Specifically, the reward function is intrinsically  tied to the average QFI, denoted as \(\bar{\mathcal{F}}_{\Omega}\) in Eq.~(\ref{aqfi}), which governs the fundamental precision limit of rotational velocity estimation. To ensure numerical stability during RL optimization, we implement a normalized reward formulation: \(r_i= \bar{\mathcal{F}}^i_{\Omega}/10^{17}\), effectively constraining the reward magnitude within a tractable scale for gradient-based policy updates.
	By optimizing the QFI, the agent learns policies that enhance the gyroscope's sensitivity.
	The  action space \(\mathcal{A}\) is defined as a continuously controllable variable, specifically \(a_t = \Delta_c(t)/\omega_m\), which is updated within the interval \([t, t + \delta \tau)\). In practice, the pump-cavity detuning \(\Delta_c(t)\) can be efficiently controlled using an arbitrary waveform generator (AWG)~\cite{Ai2022,Zhang2023,PhysRevLett.132.213602}, enabling precise and real-time adjustments to optimize the system’s performance.
	
	The PPO algorithm iteratively samples trajectories from the environment to update the policy \(\pi(a|s)\) and the value function \(Q(s, a)\). At each training step, the agent collects a batch of experiences \(\{ (s_t, a_t, r_{t+1}, s_{t+1}) \}\) and computes the advantage function \(A(s_t, a_t)\), defined as:
   \(A(s_t, a_t) = Q(s_t, a_t) - V(s_t),\)
	where \(Q(s_t, a_t)\) is the action-value function and \(V(s_t)\) is the state-value function. The advantage function quantifies the relative value of action \(a_t\) compared to the average action in state \(s_t\). The policy is updated by maximizing a clipped objective function:
    \(	L^{\text{CLIP}}(\theta) = \mathbb{E}_t \left[ \min \left( {\bf p}_t(\theta) A(s_t, a_t), \text{clip}({\bf p}_t(\theta), 1 - \epsilon, 1 + \epsilon) A(s_t, a_t) \right) \right],\)
	where \({\bf{p}}_t(\theta) = \frac{\pi_\theta(a_t|s_t)}{\pi_{\theta_{\text{old}}}(a_t|s_t)}\) is the probability ratio between the new and old policies, and \(\epsilon\) (typically 0.2) controls the clipping range. This mechanism prevents large policy deviations while maintaining a balance between exploration and exploitation.
	The value function \(V(s)\) is updated by minimizing the mean squared error between predicted and target values:
    \(L^{\text{VF}}(\theta) = \mathbb{E}_t \left[ \left( V_\theta(s_t) - V^{\text{target}}_t \right)^2\right],\)
	where \(V^{\text{target}}_t\) is computed using the Bellman equation. The total PPO objective combines the clipped policy objective, value function objective, and an entropy bonus \(S[\pi_\theta](s_t)\) to encourage exploration:
    \(	L^{\text{PPO}}(\theta) = L^{\text{CLIP}}(\theta) - c_1 L^{\text{VF}}(\theta) + c_2 S[\pi_\theta](s_t),\)
	where \(c_1\) and \(c_2\) are weighting coefficients~\cite{PPO}. This approach enables the agent to effectively navigate the dynamics of complex systems, as demonstrated in Fig.~\ref{Fig4}(a).

    By leveraging the PPO algorithm, we achieve adaptive control of the pump-cavity detuning \(\Delta_c(t)/\omega_m\), optimizing the average QFI \(\bar{\mathcal{F}}_{\Omega}\)  and  estimation errors \(\Delta \Omega\) in the presence of environmental noise and thermal fluctuations. This adaptive control framework enables the system to dynamically adjust to varying operational conditions, ensuring optimal performance across a wide range of parameters while effectively overcoming the influence of non-reciprocity on the direction of angular velocity.

    In Figs.~\ref{Fig4}(b-d), we employ the PPO algorithm to determine the optimal pump-cavity detuning \(\Delta_c\) for various values of the parameter \(\varepsilon\). We investigate its influence on the mean QFI, \(\bar{\mathcal{F}}_{\Omega}\), and the corresponding mean frequency estimation precision, \(\Delta\bar{\Omega}\), across the angular velocity range \(\Omega \in [-4, 4]\)kHz, which spans the effective operational range of the non-reciprocal gyroscope.
	The entire dynamical evolution is discretized into 10 time steps, with each interval set to \(\delta\tau = 2/\omega_m\), ensuring a fine-grained optimization process.
	Figure \ref{Fig4}(b) illustrates an example of the {optimal control sequence} \(\Delta_c/\omega_m \in [-1.5, 1.5]\) obtained through the PPO algorithm. This sequence represents the dynamically adjusted detuning values that maximize the system's sensitivity. Using this optimized sequence, we analyze the resulting dynamical behavior of \(\bar{\mathcal{F}}_{\Omega}\) and \(\Delta\bar{\Omega}\) for various values of \(\varepsilon\) across the frequency range \(\Omega \in [-4, 4]\)kHz. As shown in Fig.~\ref{Fig4}(c), \(\bar{\mathcal{F}}_{\Omega}\) increases monotonically with \(\varepsilon\), indicating enhanced sensitivity of the gyroscope at higher \(\varepsilon\) values. This trend demonstrates the effectiveness of RL in tailoring the system's response to achieve superior sensitivity.
	Furthermore, Fig.~\ref{Fig4}(d) presents the corresponding evolution of the {mean estimation precision} \(\Delta\bar{\Omega}\). The results reveal a significant improvement in precision, underscoring the ability of the PPO algorithm to minimize estimation errors and enhance the reliability of frequency measurements.
    A comparative analysis of Figs.~\ref{Fig2} and \ref{Fig3} demonstrates that RL enables the achievement of average QFI and average estimation error metrics within the specified angular velocity measurement range, matching the precision scale of optimal local angular velocity measurement schemes.

\begin{figure*}[tbp]
	\centering\includegraphics[width=12cm]{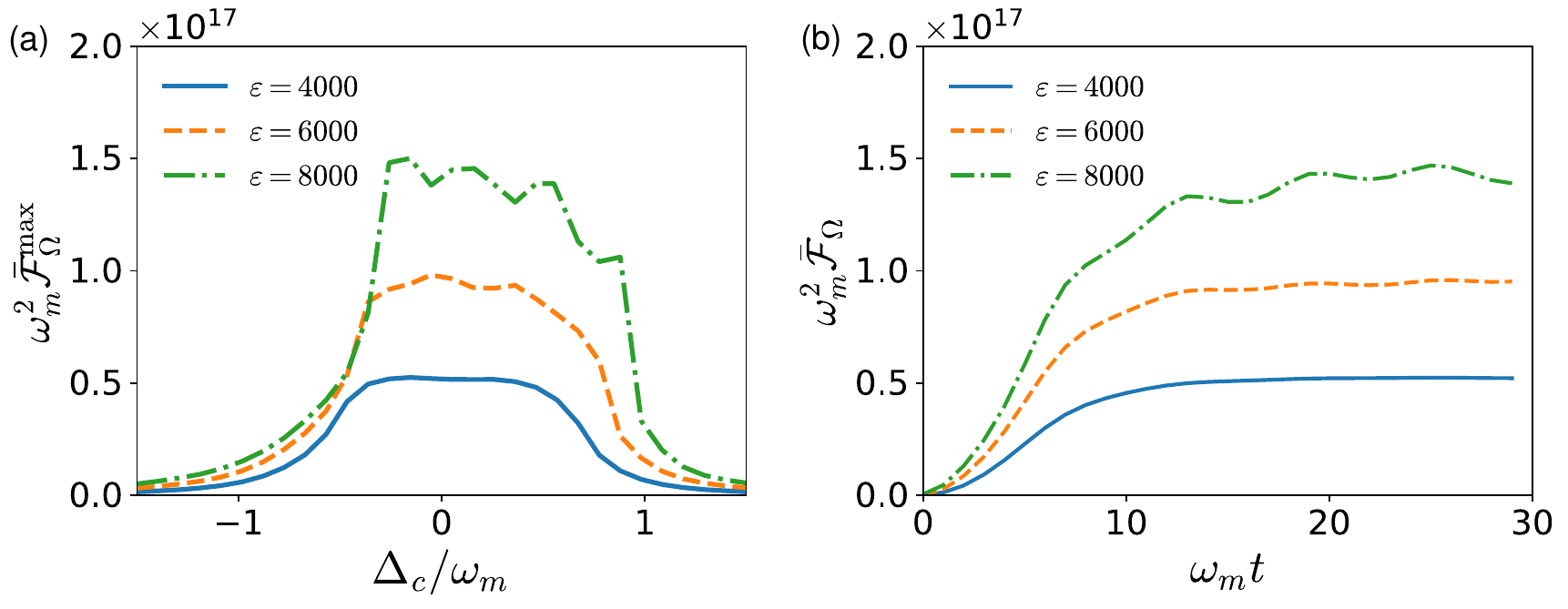}
	\caption{ (a) Dependence of the optimal average QFI $\bar{\mathcal{F}}^{\rm max}_{\Omega}$ on the normalized parameter $\Delta_c/\omega_m$ under fixed action values.
		(b) Time evolution of the average QFI $\bar{\mathcal{F}}_{\Omega}(t)$ with optimally fixed parameter $\Delta_c/\omega_m=-0.1$. All other parameters are consistent with those used in Fig.~\ref{Fig4}.
	} \label{Fig5}
\end{figure*}
    To further demonstrate the effectiveness of the RL approach as shown in Fig.~\ref{Fig4}, Fig.~\ref{Fig5}(a) presents the optimal average QFI $\bar{\mathcal{F}}^{\rm max}_{\Omega}$ obtained through a numerical search over time under fixed action parameters $\Delta_c/\omega_m$. Figure~\ref{Fig5}(b) shows the corresponding dynamic evolution of the average QFI when $\Delta_c/\omega_m$ is fixed at its optimal value of approximately $-0.1$. It is observed that as $\omega_m t \to 20$, the average QFI $\bar{\mathcal{F}}_{\Omega}(t)$ approaches a steady value, although small oscillations persist for large $\varepsilon$. 
   In contrast, the RL-based optimization shown in Fig.~\ref{Fig4} easily surpasses the fixed-detuning optimum presented in Fig.~\ref{Fig5}, achieving superior performance within just 20 time steps $ \omega_m t = 20$. This underscores the efficiency of the RL approach in rapidly guiding the system toward optimal sensitivity.

    Based on the comparison between Figs.~\ref{Fig4} and \ref{Fig5}, we highlight the advantage of RL-based dynamic detuning. In Fig.~\ref{Fig4}(b), the detuning is adaptively adjusted to optimize the average QFI. As the evolution progresses toward $\omega_m t \to 20$, the $\bar{\mathcal{F}}_{\Omega}(t)$  approaches its optimal value, with the detuning of the last step  approximately $-1.0$ at $\omega_m t = 20$. 
    This adaptive adjustment suggests that, after reaching near-optimal conditions, the control strategy tends to suppress excessive system evolution, which would otherwise diminish the average QFI $\bar{\mathcal{F}}_{\Omega}(t) $ accumulated during earlier stages. As shown in Fig.~\ref{Fig5},  maintaining a large fixed detuning, such as  $\Delta_c/\omega_m \approx -1.0 $, suppresses the system's evolution and leads to a significant reduction in sensing performance.

    The RL-based strategy differs fundamentally from fixed-parameter approaches by continuously adapting the control in response to the system’s instantaneous quantum state. It actively shapes the entire dynamical trajectory, enabling more efficient and globally optimal average QFI accumulation. The comparison with static optimization strategies clearly illustrates that RL-guided control consistently outperforms even the best fixed settings, underscoring its ability to autonomously learn and implement high-performance control policies in optomechanical gyroscopes.

    Moreover, by adaptively tuning the system parameters in real time, the RL framework not only enhances the precision of frequency estimation, but also improves robustness against parameter fluctuations. This adaptability directly addresses key limitations in traditional local optomechanical gyroscopes, marking a significant step toward high-precision, resilient quantum sensing.

	\section{Discussion and conclusion}\label{VI}
    In summary, our innovative optomechanical gyroscope surpasses the SQL in angular velocity estimation without the need for direct utilization of quantum light sources. By integrating a spinning COM resonator with a tapered fiber, we precisely determine the resonator's angular velocity through measurements of the output optical field. Our analysis, incorporating quantum QFI calculations and optimal angular velocity estimation precision, reveals significant quantum non-reciprocity effects influenced by the propagation direction of the driving optical field. Strategic adjustments in driving directions and intensities further enhance the gyroscope's sensitivity.
    To address the limitation of localized angular velocity sensitivity, we propose leveraging RL techniques to broaden sensitivity across a wider range of angular velocities. By optimizing the detuning between pump light and spinning cavity frequencies using RL, the system autonomously learns through reward function feedback, extending sensitivity to angular velocities over a broader operational range.

    Our optomechanical gyroscope achieves high-precision rotational velocity sensing in a compact and integrable configuration. Furthermore, RL techniques hold significant promise for further enhancing gyroscope performance, driving progress in quantum sensing technology. The investigation of non-reciprocal properties in such gyroscopes and their impact on rotation angular velocity estimation accuracy is crucial for advancing their performance and applications. This research has the potential to revolutionize diverse fields, including aerospace engineering, robotics, and fundamental physics experiments requiring precise rotation sensing.

	\acknowledgments
	Q.S.T. acknowledges support from the National
	Natural Science Foundation of China (NSFC) (12275077).
	Y.-F.J. is supported by the NSFC (Grant No. 12405029) and the Natural Science Foundation of Henan Province (Grant No. 252300421221).
    L.X. acknowledges support from the Natural Science Foundation of Hunan   Province (Grant No. 2024JJ5102) and the Key Laboratory of Low-Dimensional Quantum Structures and Quantum Control of Ministry of Education (Grant No.QSQC2405).
	J.-Q.L. was supported in part by National Natural
	Science Foundation of China (Grants No. 12175061,
	No. 12247105, No. 11935006, and No. 12421005), National Key Research and Development Program of China
	(Grant No. 2024YFE0102400), and Hunan Provincial
	Major Sci-Tech Program (Grant No. 2023ZJ1010).
  L.M.K. is supported by the Innovation Program for Quantum Science and Technology (Grant No. 2024ZD0301000), the NSFC (Grant Nos.  12247105, 12175060, 12421005), and the XJ-Lab key project (Grant No.23XJ02001).
Y.Z. acknowledges support from
the Natural Science Foundation of Hunan Province (Grant
No. 2025JJ60018) and the Scientific Research Foundation
of the Hunan Provincial Education Department (Grant No.
24B0866).

	\appendix
	\begin{widetext}
	\section{The equation of motion for the second-order moments} \label{appa}
	
	In this appendix, we present the matrix formulation for solving the dynamics of second-order moments as described in Eq.~(\ref{xx}) of the main text.
	
		\begin{equation}
			\resizebox{1.05\textwidth}{!}{
				$
				\mathbf{A} = \left[ 
				\begin{array}{ccccccccccccccccccccc}
					-\kappa  & 0 & 0 & 0 & 0 & 0 & 0 & 0 & 0 & -iJ & iJ & 0 & 0 
					& 
					iG_{\circlearrowleft} & 
					-iG_{\circlearrowleft}^{\ast } & iG_{\circlearrowleft} & 
					-iG_{\circlearrowleft}^{\ast } & 0 & 0 & 0 & 0 \\ 
					0 & -\kappa  & 0 & 0 & 0 & 0 & 0 & 0 & 0 & iJ & -iJ & 0 & 0 
					& 0 
					& 0 & 0 & 0
					& iG_{\circlearrowright} & -iG_{\circlearrowright}^{\ast } 
					& 
					iG_{\circlearrowright} & -iG_{\circlearrowright}^{\ast } \\ 
					0 & 0 & -\gamma _{m} & 0 & 0 & 0 & 0 & 0 & 0 & 0 & 0 & 0 & 
					0 & 
					-iG_{\circlearrowleft} & 
					iG_{\circlearrowleft}^{\ast } & iG_{\circlearrowleft} & 
					-iG_{\circlearrowleft}^{\ast } & -iG_{\circlearrowright} & 
					iG_{\circlearrowright}^{\ast } & iG_{\circlearrowright}
					& -iG_{\circlearrowright}^{\ast } \\ 
					0 & 0 & 0 & h_{3} & 0 & 0 & 0 & 0 & 0 & 0 & 0 & 2iJ & 0 & 
					-2iG_{\circlearrowleft}^{\ast } & 
					0 & -2iG_{\circlearrowleft}^{\ast } & 0 & 0 & 0 & 0 & 0 \\ 
					0 & 0 & 0 & 0 & h_{4} & 0 & 0 & 0 & 0 & 0 & 0 & 0 & -2iJ & 
					0 & 
					2iG_{\circlearrowleft} & 0 & 
					2iG_{\circlearrowleft} & 0 & 0 & 0 & 0 \\ 
					0 & 0 & 0 & 0 & 0 & h_{5} & 0 & 0 & 0 & 0 & 0 & 2iJ & 0 & 0 
					& 0 
					& 0 & 0 & 
					-2iG_{\circlearrowright}^{\ast } & 0 & 
					-2iG_{\circlearrowright}^{\ast } & 0 \\ 
					0 & 0 & 0 & 0 & 0 & 0 & h_{6} & 0 & 0 & 0 & 0 & 0 & -2iJ & 
					0 & 
					0 & 0 & 0 & 0
					& 2iG_{\circlearrowright} & 0 & 2iG_{\circlearrowright} \\ 
					0 & 0 & 0 & 0 & 0 & 0 & 0 & h_{7} & 0 & 0 & 0 & 0 & 0 & 0 & 
					-2iG_{\circlearrowleft}^{\ast }
					& -2iG_{\circlearrowleft} & 0 & 0 & 
					-2iG_{\circlearrowright}^{\ast } & -2iG_{\circlearrowright} 
					& 0 \\ 
					0 & 0 & 0 & 0 & 0 & 0 & 0 & 0 & h_{8} & 0 & 0 & 0 & 0 & 
					2iG_{\circlearrowleft} 
					& 0 & 0 & 
					2iG_{\circlearrowleft}^{\ast } & 2iG_{\circlearrowright} & 
					0 & 
					0 & 
					2iG_{\circlearrowright}^{\ast } \\ 
					-iJ & iJ & 0 & 0 & 0 & 0 & 0 & 0 & 0 & h_{9} & 0 & 0 & 0 & 
					iG_{\circlearrowright} & 0 & 
					iG_{\circlearrowright} & 0 & 0 & 
					-iG_{\circlearrowleft}^{\ast } 
					& 0 & 
					-iG_{\circlearrowleft}^{\ast } \\ 
					iJ & -iJ & 0 & 0 & 0 & 0 & 0 & 0 & 0 & 0 & h_{10} & 0 & 0 & 
					0 & 
					-iG_{\circlearrowright}^{\ast } & 0 & 
					-iG_{\circlearrowright}^{\ast } & iG_{\circlearrowleft} & 0 
					& 
					iG_{\circlearrowleft} & 0 
					\\ 
					0 & 0 & 0 & iJ & 0 & iJ & 0 & 0 & 0 & 0 & 0 & h_{11} & 0 & 
					-iG_{\circlearrowright}^{\ast } & 
					0 & -iG_{\circlearrowright}^{\ast } & 0 & 
					-iG_{\circlearrowleft}^{\ast } & 0 & 
					-iG_{\circlearrowleft}^{\ast } 
					& 0 \\ 
					0 & 0 & 0 & 0 & -iJ & 0 & -iJ & 0 & 0 & 0 & 0 & 0 & h_{12} 
					& 0 
					& iG_{\circlearrowright} & 0
					& iG_{\circlearrowright} & 0 & iG_{\circlearrowleft} & 0 & 
					iG_{\circlearrowleft} \\ 
					iG_{\circlearrowleft}^{\ast } & 0 & 
					-iG_{\circlearrowleft}^{\ast } & iG_{\circlearrowleft} & 0 
					& 0 
					& 0 & 0 & 
					-iG_{\circlearrowleft}^{\ast } & 
					iG_{\circlearrowright}^{\ast } 
					& 0 & iG_{\circlearrowright} & 
					0 & h_{13} & 0 
					& 0 & 0 & iJ
					& 0 & 0 & 0 \\ 
					-iG_{\circlearrowleft} & 0 & iG_{\circlearrowleft} & 0 & 
					-iG_{\circlearrowleft}^{\ast } & 0 & 0 & 
					iG_{\circlearrowleft} 
					& 0 
					& 0 & -iG_{\circlearrowright}
					& 0 & -iG_{\circlearrowright}^{\ast } & 0 & h_{14} & 0 & 0 
					& 0 
					& -iJ & 0 & 0 \\ 
					-iG_{\circlearrowleft}^{\ast } & 0 & 
					-iG_{\circlearrowleft}^{\ast } & -iG_{\circlearrowleft} & 0 
					& 0 
					& 0 & 
					-iG_{\circlearrowleft}^{\ast }
					& 0 & -iG_{\circlearrowright}^{\ast } & 0 & 
					-iG_{\circlearrowright} & 0 & 0 & 0 & h_{15} & 0 & 
					0 & 0 & iJ & 0
					\\ 
					iG_{\circlearrowleft} & 0 & iG_{\circlearrowleft} & 0 & 
					iG_{\circlearrowleft}^{\ast } & 0 & 0 & 0 & 
					iG_{\circlearrowleft} & 
					0 & iG_{\circlearrowright} & 
					0 & iG_{\circlearrowright}^{\ast } & 0 & 0 & 0 & h_{16} & 0 
					& 0 
					& 0 & -iJ \\ 
					0 & iG_{\circlearrowright}^{\ast } & 
					-iG_{\circlearrowright}^{\ast } & 0 & 0 & 
					iG_{\circlearrowright} & 0 & 0 & 
					-iG_{\circlearrowright}^{\ast } & 0 & 
					iG_{\circlearrowleft}^{\ast } & 
					iG_{\circlearrowleft} & 0 & iJ & 0 & 0 
					& 0 & h_{17}
					& 0 & 0 & 0 \\ 
					0 & -iG_{\circlearrowright} & iG_{\circlearrowright} & 0 & 
					0 & 
					0 & -iG_{\circlearrowright}^{\ast } & 
					iG_{\circlearrowright} & 
					0 
					& -iG_{\circlearrowleft} & 0
					& 0 & -iG_{\circlearrowleft}^{\ast } & 0 & -iJ & 0 & 0 & 0 
					& 
					h_{18} & 0 & 0 \\ 
					0 & -iG_{\circlearrowright}^{\ast } & 
					-iG_{\circlearrowright}^{\ast } & 0 & 0 & 
					-iG_{\circlearrowright} & 0 & 
					-iG_{\circlearrowright}^{\ast }
					& 0 & 0 & -iG_{\circlearrowleft}^{\ast } & 
					-iG_{\circlearrowleft} & 0 & 0 & 0 & iJ & 0 & 0 & 
					0 & h_{19} & 0
					\\ 
					0 & iG_{\circlearrowright} & iG_{\circlearrowright} & 0 & 0 
					& 0 
					& iG_{\circlearrowright}^{\ast } & 0 & 
					iG_{\circlearrowright} & 
					iG_{\circlearrowleft} & 0 & 
					0 & iG_{\circlearrowleft}^{\ast } & 0 & 0 & 0 & -iJ & 0 & 0 
					& 0 
					& h_{20}%
				\end{array}
				\right],
				$
			}\nonumber\\
		\end{equation}

		\begin{eqnarray}
			\mathbf{X}&=&[\langle 
			a_{\circlearrowleft}^{\dagger}a_{\circlearrowleft}\rangle, \langle 
			a_{\circlearrowright}^{\dagger}a_{\circlearrowright}\rangle, 
			\langle 
			b^{\dagger}b\rangle, \langle 
			a_{\circlearrowleft}^{\dagger}a_{\circlearrowleft}^{\dagger}\rangle,
			\langle a_{\circlearrowleft} a_{\circlearrowleft}\rangle, 
			\langle 
			a_{\circlearrowright}^{\dagger}a_{\circlearrowright}^{\dagger}\rangle,
			\langle a_{\circlearrowright} a_{\circlearrowright} \rangle,
			\langle b^{\dagger}b^{\dagger}\rangle, \langle 
			bb\rangle,  \langle 
			a_{\circlearrowleft}^{\dagger}a_{\circlearrowright}\rangle, \langle 
			a_{\circlearrowright}^{\dagger}a_{\circlearrowleft}\rangle, 
			\langle 
			a_{\circlearrowleft}^{\dagger}a_{\circlearrowright}^{\dagger}\rangle,
			\nonumber\\
			&&\langle a_{\circlearrowleft} a_{\circlearrowright}\rangle, 
			\langle a_{\circlearrowleft}^{\dagger}b\rangle, 
			\langle a_{\circlearrowleft} 
			b^{\dagger}\rangle, \langle 
			a_{\circlearrowleft}^{\dagger}b^{\dagger}\rangle, \langle 
			a_{\circlearrowleft} 
			b\rangle, \langle a_{\circlearrowright}^{\dagger}b\rangle, 
			\langle a_{\circlearrowright} b^{\dagger}\rangle, \langle 
			a_{\circlearrowleft}^{\dagger}b^{\dagger}\rangle, \langle 
			a_{\circlearrowright} b\rangle]^{\rm T}, 
		\end{eqnarray}

		In matrix $\bf A$, we have defined
		\begin{eqnarray*}
			&&h_3 = 2i\tilde{\Delta}_{\circlearrowleft}-\kappa, \hspace{0.2cm}	
			h_4 
			= 
			-(2i\tilde{\Delta}_{\circlearrowleft}+\kappa), \hspace{0.2cm}
			h_5 = 2i\tilde{\Delta}_{\circlearrowright}-\kappa, \hspace{0.2cm}	
			h_6 
			= 
			-(2i\tilde{\Delta}_{\circlearrowright}+\kappa)\\
			&& h_7 = 2i\omega_m -\gamma_m,\hspace{0.2cm} h_8 =-(2i\omega_m + 
			\gamma_m),
			h_9=i(\tilde{\Delta}_{\circlearrowleft}-\tilde{\Delta}_{\circlearrowright})
			-\kappa, \hspace{0.2cm}
			h_{10} 
			=-i(\tilde{\Delta}_{\circlearrowleft}-\tilde{\Delta}_{\circlearrowright})
			-\kappa\\
			&& 
			h_{11}=i(\tilde{\Delta}_{\circlearrowleft}+\tilde{\Delta}_{\circlearrowright})
			-\kappa, \hspace{0.2cm}
			h_{12} 
			=-i(\tilde{\Delta}_{\circlearrowleft}+\tilde{\Delta}_{\circlearrowright})
			-\kappa\\
			&& h_{13} = 
			[i(\tilde{\Delta}_{\circlearrowleft}-\omega_m)-\frac{\kappa+\gamma_m}{2}],
			h_{14} = 
			-[i(\tilde{\Delta}_{\circlearrowleft}-\omega_m)+\frac{\kappa+\gamma_m}{2}],\hspace{0.2cm}
			h_{15} = 
			[i(\tilde{\Delta}_{\circlearrowleft}+\omega_m)-\frac{\kappa+\gamma_m}{2}],\\
			&&h_{16} = 
			-[i(\tilde{\Delta}_{\circlearrowleft}+\omega_m)+\frac{\kappa+\gamma_m}{2}],\hspace{0.2cm}
			h_{17} = 
			[i(\tilde{\Delta}_{\circlearrowright}-\omega_m)-\frac{\kappa+\gamma_m}{2}],\hspace{0.2cm}
			h_{18} = 
			-[i(\tilde{\Delta}_{\circlearrowright}-\omega_m)+\frac{\kappa+\gamma_m}{2}],\\
			&& h_{19} = 
			[i(\tilde{\Delta}_{\circlearrowright}+\omega_m)-\frac{\kappa+\gamma_m}{2}],\hspace{0.2cm}
			h_{20} = 
			-[i(\tilde{\Delta}_{\circlearrowright}+\omega_m)+\frac{\kappa+\gamma_m}{2}].
		\end{eqnarray*}
		The inhomogeneous term $\mathbf{D}(t)$ depends on 
		$G_{\circlearrowleft}(t)$ and $G_{\circlearrowright}(t)$, and is 
		expressed 
		as:
		\begin{equation}
			\mathbf{D}(t)=[0, 0, 
			\gamma_m\bar{n}_m, 0,0, 
			0, 0, 
			0,0,0,0,0,0,0,0,-iG_{\circlearrowleft}^{*}(t), 
			iG_{\circlearrowleft}(t), 0, 0, -iG_{\circlearrowright}^{*}(t), 
			iG_{\circlearrowright}(t)].
		\end{equation}

		\section{Input-output relation} \label{appb}
		
		Treating the optical modes at different frequencies $\omega'$
		with in the dissipation  channels( $B_j$, $j=\circlearrowleft, 
		\circlearrowright $, respectively) as the ones forming an effective
		bath of the system, we can describe the complete dynamics by
		specifying apart from the system effective Hamiltonian,$H_{\rm eff}$, 
		the 
		ones
		of the bath,$H_B$,as well as the system-bath interaction, $H_{SB}$,
		in the continuum limit.
		
		\begin{eqnarray}
			H_{S,\text{eff}}& =& \omega_m b^{\dagger}b + \sum_{j= 
				\circlearrowleft, 
				\circlearrowright}\tilde{\Delta}_j 
			a_j^{\dagger}a_j - (G_j^{\ast}a_j + 
			G_ja_j^{\dagger})(b^{\dagger}+b) + 
			J(a_{\circlearrowright}^{\dagger}a_{\circlearrowleft} + 
			a_{\circlearrowleft}^{\dagger}a_{\circlearrowright}).\\
			H_B &=& \int_{-\infty}^{\infty} d\omega' \omega' 
			[B^{\dagger}_{\circlearrowleft\omega'}B_{\circlearrowleft\omega'} + 
			B^{\dagger}_{\circlearrowright\omega'}B_{\circlearrowright\omega'}],\\
			H_{SB} &=& -i\int_{-\infty}^{\infty} d\omega' \left(
			\sqrt{\frac{\kappa(\omega')}{2 \pi}}
			\left[B^{\dagger}_{\circlearrowleft\omega'}a_{\circlearrowleft} - 
			B_{\circlearrowleft\omega'}a^{\dagger}_{\circlearrowleft}\right] +
			\sqrt{\frac{\kappa(\omega')}{2\pi}}
			\left[B^{\dagger}_{\circlearrowright\omega'}a_{\circlearrowright} 
			-B_{\circlearrowright\omega'}a^{\dagger}_{\circlearrowright} \right]
			\right),
		\end{eqnarray}
		with each optical mode satisfying for any $\omega', \omega''$:
		\begin{equation}
			[ B_{\circlearrowleft\omega'}, 
			B^{\dagger}_{\circlearrowleft\omega''}] =
			[ B_{\circlearrowright\omega'}, 
			B^{\dagger}_{\circlearrowright\omega''}]=\delta(\omega'-\omega'').
		\end{equation}
		The Heisenberg equation of motion for $B_{j, \omega'}$, in the 
		interaction 
		picture, is
		\begin{eqnarray}
			\dot{B}_{j,\omega'} = -i\omega'{B}_{j,\omega'} 
			+\sqrt{\frac{\kappa(\omega')}{2 \pi}} a_j.
		\end{eqnarray}
		The solution to this equation can be written in two ways depending on 
		weather
		we choose to solve in terms of the initial conditions at time $t_0<t$ 
		(the 
		input) or
		in terms of the final conditions at times $t_1>t$, (the output). The 
		two 
		solutions are respectively
		\begin{eqnarray}\label{ini1}
			B_{j, \omega'}(t) = e^{-i\omega'(t-t_0)}B_{j, \omega'}(t_0) + 
			\sqrt{\frac{\kappa(\omega')}{2\pi}}\int_{t_0}^{t}e^{-i\omega'(t-t')}a(t')dt'
		\end{eqnarray} when $t_0<t$,
		and
		\begin{eqnarray}\label{fin1}
			B_{j, \omega'}(t) = e^{-i\omega'(t-t_1)}B_{j, \omega'}(t_1) - 
			\sqrt{\frac{\kappa(\omega')}{2\pi}}\int_{t}^{t_1}e^{-i\omega'(t-t')}a(t')dt'
		\end{eqnarray}
		when $t<t_1$.
		In physical terms $B_{j, \omega'}(t_0)$ and $B_{j, \omega'}(t_1)$ are 
		usually
		specified at $-\infty$ and $+\infty$ respectively, that is, for times 
		such 
		that the field is simply a free field, however here we only require 
		$t_0<t<t_1$.
		
		Then we can defined 
		\begin{eqnarray}
			a_{j, {\rm in}}(t) &\equiv& -\frac{1}{\sqrt{2 \pi}} \int d \omega' 
			e^{-i\omega'(t-t_0)}B_{j, \omega'}(t_0) 
		\end{eqnarray}
		with $[a_{j, {\rm in}}(t), a^{\dagger}_{j, {\rm in}}(t')]= \delta(t-t')$
		and 
		\begin{eqnarray}
			a_{j, {\rm out}}(t) &\equiv& \frac{1}{\sqrt{2 \pi}} \int d \omega' 
			e^{-i\omega'(t-t_1)}B_{j, \omega'}(t_1).
		\end{eqnarray}
		
		Assuming the Markov approximation to be $\kappa({\omega'})\approx 
		\kappa$, substituting Eqs.~(B8 and (B9) into Eqs.~(B6) and (B7) , 
		respectively, we can also obtain the following relationships
		\begin{eqnarray}
			a_{j, {\rm in}}(t) &=&\frac{\sqrt{\kappa}}{2} a(t) 
			-\frac{1}{\sqrt{2 \pi}}\int d\omega' B_{j, \omega'}(t),\\
			a_{j, {\rm out}}(t) &=& \frac{\sqrt{\kappa}}{2} a(t) 
			+\frac{1}{\sqrt{2 \pi}}\int d\omega' B_{j, \omega'}(t).
		\end{eqnarray}

		Then, the input and output fields are then seen to be related by
		\begin{equation}\label{inout}
			a_{j,  \rm out}(t) =\sqrt{\kappa}  a_j(t) -a_{j,\rm in}(t).
		\end{equation}
		The commutator for the output field may now be calculated to be
		$[a_{j, \rm out}, a^{\dagger}_{j, \rm out}]= [a_{j, \rm in}, 
		a^{\dagger}_{j, \rm in}]$ as required.
		
     	It is now possible to express the variances of the output field entirely in terms of those of the internal system:
    	\begin{equation}
		\langle a^{\dagger}_{j,  \rm out}(t), a_{j,  \rm out}(t) \rangle
		={\kappa}  \langle a^{\dagger}_{j}(t), a_{j}(t) \rangle + \langle
		a^{\dagger}_{j,  \rm in}(t), a_{j,  \rm in}(t) \rangle
		-\sqrt{\kappa}\langle a^{\dagger}_{j}(t), a_{j,  \rm in}(t) \rangle
		-\sqrt{\kappa}\langle a^{\dagger}_{j,  \rm in}(t), a_{j}(t) \rangle,
     	\end{equation}
 	    where \(\langle U, V\rangle \equiv \langle UV\rangle  -\langle U\rangle \langle V\rangle\). For the case of vacuum or coherent state inputs, the expression simplifies directly to
	    \( 	\langle a^{\dagger}_{j,  \rm out}(t), a_{j,  \rm out}(t) \rangle
	   ={\kappa}  \langle a^{\dagger}_{j}(t), a_{j}(t) \rangle \).
	   \end{widetext}
	
	\section{Experimental measurement of off-diagonal correlations} \label{appc}
	
	\begin{figure}[bp]
		\centering\includegraphics[width=8.8cm]{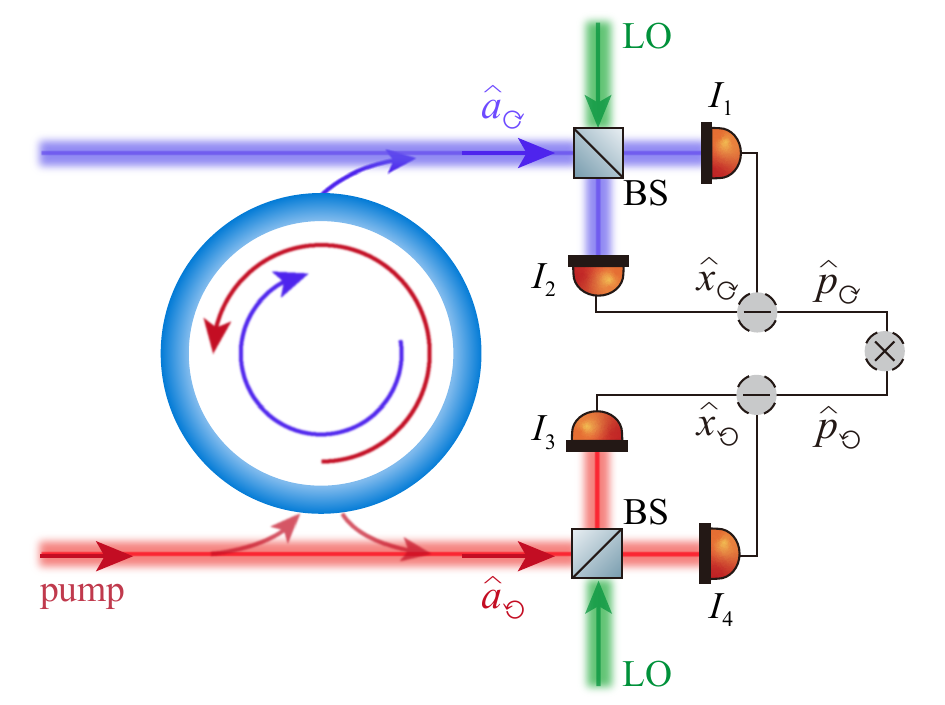}
		\caption{ Experimental protocol for quantifying off-diagonal correlations through balanced homodyne detection.	} \label{Fig6}
	\end{figure}
	To measure the off-diagonal correlations \(\langle a_{\circlearrowright}^\dagger a_{\circlearrowleft} \rangle\) and \(\langle a_{\circlearrowright} a_{\circlearrowleft} \rangle\) between optical modes \(a_{\circlearrowright}\) and \(a_{\circlearrowleft}\) as described in Eq.~(13), we employ a balanced homodyne detection scheme. The output modes \(a_{\circlearrowright,\text{out}}\) and \(a_{\circlearrowleft,\text{out}}\) are directed to separate balanced homodyne detectors, each utilizing a 50:50 beam splitter (BS) to interfere the signal mode \(a_{i,\text{out}}\) (\(i = \circlearrowright, \circlearrowleft\)) with a strong local oscillator (LO) \(a_{\text{LO}} = |a_{\text{LO}}| e^{i\theta}\). The local oscillator is matched in frequency and spatial mode to the signal, as illustrated in Fig.~\ref{Fig6}.
	
	The beam splitter outputs are detected by pairs of photodiodes, measuring the photon numbers \(I_1 = c_1^\dagger c_1\), \(I_2 = c_2^\dagger c_2\), \(I_3 = c_3^\dagger c_3\), and \(I_4 = c_4^\dagger c_4\), where the output modes are:
	\begin{eqnarray}
		c_1 = \frac{a_{\circlearrowright} + i a_{\text{LO}}}{\sqrt{2}}, \quad c_2 = \frac{  a_{\text{LO}}+ ia_{\circlearrowright}}{\sqrt{2}}, \\
		c_3 = \frac{a_{\circlearrowleft} + i a_{\text{LO}}}{\sqrt{2}}, \quad c_4 = \frac{ a_{\text{LO}} + ia_{\circlearrowleft} }{\sqrt{2}}.
	\end{eqnarray}
	
	A differential amplifier computes the differences \(I_1 - I_2\) and \(I_3 - I_4\), yielding the quadrature operators:
	\begin{eqnarray}
		I_1 - I_2 &=& 2 |a_{\text{LO}}| \left( x_{\circlearrowright} \cos\theta + p_{\circlearrowright} \sin\theta \right), \\
		I_3 - I_4 &=& 2 |a_{\text{LO}}| \left( x_{\circlearrowleft} \cos\theta + p_{\circlearrowleft} \sin\theta \right),
	\end{eqnarray}
	where the field quadratures are defined as \(x_i = (a_i + a_i^\dagger)/\sqrt{2}\) and \(p_i = (a_i - a_i^\dagger)/(i\sqrt{2})\). By setting the local oscillator phase \(\theta = 0\), we measure the \(\hat{x}\)-quadrature, while \(\theta = \pi/2\) yields the \(\hat{p}\)-quadrature.
	
	To reconstruct the off-diagonal correlations, we synchronously record the quadratures \(x_{\circlearrowright}(t_k)\), \(p_{\circlearrowright}(t_k)\), \(x_{\circlearrowleft}(t_k)\), and \(p_{\circlearrowleft}(t_k)\) at \(N\) time points \(t_k\) using phase-locked homodyne detectors. The second-order coherence properties are then evaluated using the Hanbury Brown-Twiss interferometric technique, computing the ensemble averages~\cite{doi:10.1126/science.abf2998}:
		\begin{subequations}
		\begin{align}
			\langle x_{\circlearrowright} x_{\circlearrowleft} \rangle &= \frac{1}{N} \sum_{k=1}^N x_{\circlearrowright}(t_k) x_{\circlearrowleft}(t_k), \\
			\langle x_{\circlearrowright} p_{\circlearrowleft} \rangle &= \frac{1}{N} \sum_{k=1}^N x_{\circlearrowright}(t_k) p_{\circlearrowleft}(t_k), \\
			\langle P_{\circlearrowright} x_{\circlearrowleft} \rangle &= \frac{1}{N} \sum_{k=1}^N p_{\circlearrowright}(t_k) x_{\circlearrowleft}(t_k), \\
			\langle p_{\circlearrowright} p_{\circlearrowleft} \rangle &= \frac{1}{N} \sum_{k=1}^N p_{\circlearrowright}(t_k) p_{\circlearrowleft}(t_k).
		\end{align}
	\end{subequations}

   For sufficiently large  $N$, these correlations provide complete characterization of the off-diagonal elements in the system's density matrix.
 The expectation values are obtained as:
	\begin{eqnarray}
		\langle a_{\circlearrowright}^\dagger a_{\circlearrowleft} \rangle &=& \frac{1}{2} \left[ \langle x_{\circlearrowright} x_{\circlearrowleft} \rangle - \langle p_{\circlearrowright} p_{\circlearrowleft} \rangle + i \left( \langle x_{\circlearrowright} p_{\circlearrowleft} \rangle + \langle p_{\circlearrowright} x_{\circlearrowleft} \rangle \right) \right], \nonumber \\
		\langle a_{\circlearrowright} a_{\circlearrowleft} \rangle &=& \frac{1}{2} \left[ \langle x_{\circlearrowright} x_{\circlearrowleft} \rangle + \langle p_{\circlearrowright} p_{\circlearrowleft} \rangle + i \left( \langle x_{\circlearrowright} p_{\circlearrowleft} \rangle - \langle p_{\circlearrowright} x_{\circlearrowleft} \rangle \right) \right].\nonumber \\
	\end{eqnarray}
	Additional correlation terms can be measured using analogous procedures. This approach provides a robust framework for characterizing the quantum coherence and entanglement between the optical modes.
	
\bibliography{reference}

\begin{thebibliography}{55}%
\makeatletter
\providecommand \@ifxundefined [1]{%
 \@ifx{#1\undefined}
}%
\providecommand \@ifnum [1]{%
 \ifnum #1\expandafter \@firstoftwo
 \else \expandafter \@secondoftwo
 \fi
}%
\providecommand \@ifx [1]{%
 \ifx #1\expandafter \@firstoftwo
 \else \expandafter \@secondoftwo
 \fi
}%
\providecommand \natexlab [1]{#1}%
\providecommand \enquote  [1]{``#1''}%
\providecommand \bibnamefont  [1]{#1}%
\providecommand \bibfnamefont [1]{#1}%
\providecommand \citenamefont [1]{#1}%
\providecommand \href@noop [0]{\@secondoftwo}%
\providecommand \href [0]{\begingroup \@sanitize@url \@href}%
\providecommand \@href[1]{\@@startlink{#1}\@@href}%
\providecommand \@@href[1]{\endgroup#1\@@endlink}%
\providecommand \@sanitize@url [0]{\catcode `\\12\catcode `\$12\catcode
  `\&12\catcode `\#12\catcode `\^12\catcode `\_12\catcode `\%12\relax}%
\providecommand \@@startlink[1]{}%
\providecommand \@@endlink[0]{}%
\providecommand \url  [0]{\begingroup\@sanitize@url \@url }%
\providecommand \@url [1]{\endgroup\@href {#1}{\urlprefix }}%
\providecommand \urlprefix  [0]{URL }%
\providecommand \Eprint [0]{\href }%
\providecommand \doibase [0]{https://doi.org/}%
\providecommand \selectlanguage [0]{\@gobble}%
\providecommand \bibinfo  [0]{\@secondoftwo}%
\providecommand \bibfield  [0]{\@secondoftwo}%
\providecommand \translation [1]{[#1]}%
\providecommand \BibitemOpen [0]{}%
\providecommand \bibitemStop [0]{}%
\providecommand \bibitemNoStop [0]{.\EOS\space}%
\providecommand \EOS [0]{\spacefactor3000\relax}%
\providecommand \BibitemShut  [1]{\csname bibitem#1\endcsname}%
\let\auto@bib@innerbib\@empty
\bibitem [{\citenamefont {Chow}\ \emph {et~al.}(1985)\citenamefont {Chow},
  \citenamefont {Gea-Banacloche}, \citenamefont {Pedrotti}, \citenamefont
  {Sanders}, \citenamefont {Schleich},\ and\ \citenamefont
  {Scully}}]{RevModPhys.57.61}%
  \BibitemOpen
  \bibfield  {author} {\bibinfo {author} {\bibfnamefont {W.~W.}\ \bibnamefont
  {Chow}}, \bibinfo {author} {\bibfnamefont {J.}~\bibnamefont
  {Gea-Banacloche}}, \bibinfo {author} {\bibfnamefont {L.~M.}\ \bibnamefont
  {Pedrotti}}, \bibinfo {author} {\bibfnamefont {V.~E.}\ \bibnamefont
  {Sanders}}, \bibinfo {author} {\bibfnamefont {W.}~\bibnamefont {Schleich}},\
  and\ \bibinfo {author} {\bibfnamefont {M.~O.}\ \bibnamefont {Scully}},\
  }\bibfield  {title} {\bibinfo {title} {The ring laser gyro},\ }\href
  {https://doi.org/10.1103/RevModPhys.57.61} {\bibfield  {journal} {\bibinfo
  {journal} {Rev. Mod. Phys.}\ }\textbf {\bibinfo {volume} {57}},\ \bibinfo
  {pages} {61} (\bibinfo {year} {1985})}\BibitemShut {NoStop}%
\bibitem [{\citenamefont {Yadav}\ \emph {et~al.}(2012)\citenamefont {Yadav},
  \citenamefont {Shimon},\ and\ \citenamefont {Keating}}]{PhysRevD.86.083002}%
  \BibitemOpen
  \bibfield  {author} {\bibinfo {author} {\bibfnamefont {A.~P.~S.}\
  \bibnamefont {Yadav}}, \bibinfo {author} {\bibfnamefont {M.}~\bibnamefont
  {Shimon}},\ and\ \bibinfo {author} {\bibfnamefont {B.~G.}\ \bibnamefont
  {Keating}},\ }\bibfield  {title} {\bibinfo {title} {Revealing cosmic
  rotation},\ }\href {https://doi.org/10.1103/PhysRevD.86.083002} {\bibfield
  {journal} {\bibinfo  {journal} {Phys. Rev. D}\ }\textbf {\bibinfo {volume}
  {86}},\ \bibinfo {pages} {083002} (\bibinfo {year} {2012})}\BibitemShut
  {NoStop}%
\bibitem [{\citenamefont {POST}(1967)}]{RevModPhys.39.475}%
  \BibitemOpen
  \bibfield  {author} {\bibinfo {author} {\bibfnamefont {E.~J.}\ \bibnamefont
  {POST}},\ }\bibfield  {title} {\bibinfo {title} {Sagnac effect},\ }\href
  {https://doi.org/10.1103/RevModPhys.39.475} {\bibfield  {journal} {\bibinfo
  {journal} {Rev. Mod. Phys.}\ }\textbf {\bibinfo {volume} {39}},\ \bibinfo
  {pages} {475} (\bibinfo {year} {1967})}\BibitemShut {NoStop}%
\bibitem [{\citenamefont {Arditty}\ and\ \citenamefont
  {Lef\`{e}vre}(1981)}]{OL1981}%
  \BibitemOpen
  \bibfield  {author} {\bibinfo {author} {\bibfnamefont {H.~J.}\ \bibnamefont
  {Arditty}}\ and\ \bibinfo {author} {\bibfnamefont {H.~C.}\ \bibnamefont
  {Lef\`{e}vre}},\ }\bibfield  {title} {\bibinfo {title} {Sagnac effect in
  fiber gyroscopes},\ }\href {https://doi.org/10.1364/OL.6.000401} {\bibfield
  {journal} {\bibinfo  {journal} {Opt. Lett.}\ }\textbf {\bibinfo {volume}
  {6}},\ \bibinfo {pages} {401} (\bibinfo {year} {1981})}\BibitemShut {NoStop}%
\bibitem [{\citenamefont {Zhang}\ \emph {et~al.}(2025)\citenamefont {Zhang},
  \citenamefont {Zhang}, \citenamefont {Zuo},\ and\ \citenamefont
  {Kuang}}]{kuang2025}%
  \BibitemOpen
  \bibfield  {author} {\bibinfo {author} {\bibfnamefont {W.-X.}\ \bibnamefont
  {Zhang}}, \bibinfo {author} {\bibfnamefont {R.}~\bibnamefont {Zhang}},
  \bibinfo {author} {\bibfnamefont {Y.}~\bibnamefont {Zuo}},\ and\ \bibinfo
  {author} {\bibfnamefont {L.-M.}\ \bibnamefont {Kuang}},\ }\bibfield  {title}
  {\bibinfo {title} {Enhancing the sensitivity of quantum fiber-optical
  gyroscope via a non-gaussian-state probe},\ }\href
  {https://doi.org/https://doi.org/10.1002/qute.202400270} {\bibfield
  {journal} {\bibinfo  {journal} {Adv. Quantum Technol.}\ }\textbf {\bibinfo
  {volume} {8}},\ \bibinfo {pages} {2400270} (\bibinfo {year}
  {2025})}\BibitemShut {NoStop}%
\bibitem [{\citenamefont {Dowling}(1998)}]{PhysRevA.57.4736}%
  \BibitemOpen
  \bibfield  {author} {\bibinfo {author} {\bibfnamefont {J.~P.}\ \bibnamefont
  {Dowling}},\ }\bibfield  {title} {\bibinfo {title} {Correlated input-port,
  matter-wave interferometer: Quantum-noise limits to the atom-laser
  gyroscope},\ }\href {https://doi.org/10.1103/PhysRevA.57.4736} {\bibfield
  {journal} {\bibinfo  {journal} {Phys. Rev. A}\ }\textbf {\bibinfo {volume}
  {57}},\ \bibinfo {pages} {4736} (\bibinfo {year} {1998})}\BibitemShut
  {NoStop}%
\bibitem [{\citenamefont {Kok}\ \emph {et~al.}(2017)\citenamefont {Kok},
  \citenamefont {Dunningham},\ and\ \citenamefont
  {Ralph}}]{PhysRevA.95.012326}%
  \BibitemOpen
  \bibfield  {author} {\bibinfo {author} {\bibfnamefont {P.}~\bibnamefont
  {Kok}}, \bibinfo {author} {\bibfnamefont {J.}~\bibnamefont {Dunningham}},\
  and\ \bibinfo {author} {\bibfnamefont {J.~F.}\ \bibnamefont {Ralph}},\
  }\bibfield  {title} {\bibinfo {title} {Role of entanglement in calibrating
  optical quantum gyroscopes},\ }\href
  {https://doi.org/10.1103/PhysRevA.95.012326} {\bibfield  {journal} {\bibinfo
  {journal} {Phys. Rev. A}\ }\textbf {\bibinfo {volume} {95}},\ \bibinfo
  {pages} {012326} (\bibinfo {year} {2017})}\BibitemShut {NoStop}%
\bibitem [{\citenamefont {Moan}\ \emph {et~al.}(2020)\citenamefont {Moan},
  \citenamefont {Horne}, \citenamefont {Arpornthip}, \citenamefont {Luo},
  \citenamefont {Fallon}, \citenamefont {Berl},\ and\ \citenamefont
  {Sackett}}]{PhysRevLett.124.120403}%
  \BibitemOpen
  \bibfield  {author} {\bibinfo {author} {\bibfnamefont {E.~R.}\ \bibnamefont
  {Moan}}, \bibinfo {author} {\bibfnamefont {R.~A.}\ \bibnamefont {Horne}},
  \bibinfo {author} {\bibfnamefont {T.}~\bibnamefont {Arpornthip}}, \bibinfo
  {author} {\bibfnamefont {Z.}~\bibnamefont {Luo}}, \bibinfo {author}
  {\bibfnamefont {A.~J.}\ \bibnamefont {Fallon}}, \bibinfo {author}
  {\bibfnamefont {S.~J.}\ \bibnamefont {Berl}},\ and\ \bibinfo {author}
  {\bibfnamefont {C.~A.}\ \bibnamefont {Sackett}},\ }\bibfield  {title}
  {\bibinfo {title} {Quantum rotation sensing with dual sagnac interferometers
  in an atom-optical waveguide},\ }\href
  {https://doi.org/10.1103/PhysRevLett.124.120403} {\bibfield  {journal}
  {\bibinfo  {journal} {Phys. Rev. Lett.}\ }\textbf {\bibinfo {volume} {124}},\
  \bibinfo {pages} {120403} (\bibinfo {year} {2020})}\BibitemShut {NoStop}%
\bibitem [{\citenamefont {Aspelmeyer}\ \emph {et~al.}(2014)\citenamefont
  {Aspelmeyer}, \citenamefont {Kippenberg},\ and\ \citenamefont
  {Marquardt}}]{RevModPhys.86.1391}%
  \BibitemOpen
  \bibfield  {author} {\bibinfo {author} {\bibfnamefont {M.}~\bibnamefont
  {Aspelmeyer}}, \bibinfo {author} {\bibfnamefont {T.~J.}\ \bibnamefont
  {Kippenberg}},\ and\ \bibinfo {author} {\bibfnamefont {F.}~\bibnamefont
  {Marquardt}},\ }\bibfield  {title} {\bibinfo {title} {Cavity optomechanics},\
  }\href {https://doi.org/10.1103/RevModPhys.86.1391} {\bibfield  {journal}
  {\bibinfo  {journal} {Rev. Mod. Phys.}\ }\textbf {\bibinfo {volume} {86}},\
  \bibinfo {pages} {1391} (\bibinfo {year} {2014})}\BibitemShut {NoStop}%
\bibitem [{\citenamefont {Verhagen}\ \emph {et~al.}(2012)\citenamefont
  {Verhagen}, \citenamefont {Del{\'e}glise}, \citenamefont {Weis},
  \citenamefont {Schliesser},\ and\ \citenamefont {Kippenberg}}]{Verhagen2012}%
  \BibitemOpen
  \bibfield  {author} {\bibinfo {author} {\bibfnamefont {E.}~\bibnamefont
  {Verhagen}}, \bibinfo {author} {\bibfnamefont {S.}~\bibnamefont
  {Del{\'e}glise}}, \bibinfo {author} {\bibfnamefont {S.}~\bibnamefont {Weis}},
  \bibinfo {author} {\bibfnamefont {A.}~\bibnamefont {Schliesser}},\ and\
  \bibinfo {author} {\bibfnamefont {T.~J.}\ \bibnamefont {Kippenberg}},\
  }\bibfield  {title} {\bibinfo {title} {Quantum-coherent coupling of a
  mechanical oscillator to an optical cavity mode},\ }\href
  {https://doi.org/10.1038/nature10787} {\bibfield  {journal} {\bibinfo
  {journal} {Nature (London)}\ }\textbf {\bibinfo {volume} {482}},\ \bibinfo
  {pages} {63} (\bibinfo {year} {2012})}\BibitemShut {NoStop}%
\bibitem [{\citenamefont {Davuluri}\ \emph {et~al.}(2017)\citenamefont
  {Davuluri}, \citenamefont {Li},\ and\ \citenamefont {Li}}]{Davuluri_2017}%
  \BibitemOpen
  \bibfield  {author} {\bibinfo {author} {\bibfnamefont {S.}~\bibnamefont
  {Davuluri}}, \bibinfo {author} {\bibfnamefont {K.}~\bibnamefont {Li}},\ and\
  \bibinfo {author} {\bibfnamefont {Y.}~\bibnamefont {Li}},\ }\bibfield
  {title} {\bibinfo {title} {Gyroscope with two-dimensional optomechanical
  mirror},\ }\href {https://doi.org/10.1088/1367-2630/aa8afb} {\bibfield
  {journal} {\bibinfo  {journal} {New J. Phys.}\ }\textbf {\bibinfo {volume}
  {19}},\ \bibinfo {pages} {113004} (\bibinfo {year} {2017})}\BibitemShut
  {NoStop}%
\bibitem [{\citenamefont {Romero-Isart}\ \emph {et~al.}(2011)\citenamefont
  {Romero-Isart}, \citenamefont {Pflanzer}, \citenamefont {Blaser},
  \citenamefont {Kaltenbaek}, \citenamefont {Kiesel}, \citenamefont
  {Aspelmeyer},\ and\ \citenamefont {Cirac}}]{PhysRevLett.107.020405}%
  \BibitemOpen
  \bibfield  {author} {\bibinfo {author} {\bibfnamefont {O.}~\bibnamefont
  {Romero-Isart}}, \bibinfo {author} {\bibfnamefont {A.~C.}\ \bibnamefont
  {Pflanzer}}, \bibinfo {author} {\bibfnamefont {F.}~\bibnamefont {Blaser}},
  \bibinfo {author} {\bibfnamefont {R.}~\bibnamefont {Kaltenbaek}}, \bibinfo
  {author} {\bibfnamefont {N.}~\bibnamefont {Kiesel}}, \bibinfo {author}
  {\bibfnamefont {M.}~\bibnamefont {Aspelmeyer}},\ and\ \bibinfo {author}
  {\bibfnamefont {J.~I.}\ \bibnamefont {Cirac}},\ }\bibfield  {title} {\bibinfo
  {title} {Large quantum superpositions and interference of massive
  nanometer-sized objects},\ }\href
  {https://doi.org/10.1103/PhysRevLett.107.020405} {\bibfield  {journal}
  {\bibinfo  {journal} {Phys. Rev. Lett.}\ }\textbf {\bibinfo {volume} {107}},\
  \bibinfo {pages} {020405} (\bibinfo {year} {2011})}\BibitemShut {NoStop}%
\bibitem [{\citenamefont {Nimmrichter}\ \emph {et~al.}(2014)\citenamefont
  {Nimmrichter}, \citenamefont {Hornberger},\ and\ \citenamefont
  {Hammerer}}]{PhysRevLett.113.020405}%
  \BibitemOpen
  \bibfield  {author} {\bibinfo {author} {\bibfnamefont {S.}~\bibnamefont
  {Nimmrichter}}, \bibinfo {author} {\bibfnamefont {K.}~\bibnamefont
  {Hornberger}},\ and\ \bibinfo {author} {\bibfnamefont {K.}~\bibnamefont
  {Hammerer}},\ }\bibfield  {title} {\bibinfo {title} {Optomechanical sensing
  of spontaneous wave-function collapse},\ }\href
  {https://doi.org/10.1103/PhysRevLett.113.020405} {\bibfield  {journal}
  {\bibinfo  {journal} {Phys. Rev. Lett.}\ }\textbf {\bibinfo {volume} {113}},\
  \bibinfo {pages} {020405} (\bibinfo {year} {2014})}\BibitemShut {NoStop}%
\bibitem [{\citenamefont {Massel}\ \emph {et~al.}(2012)\citenamefont {Massel},
  \citenamefont {Cho}, \citenamefont {Pirkkalainen}, \citenamefont {Hakonen},
  \citenamefont {Heikkil{\"a}},\ and\ \citenamefont
  {Sillanp{\"a}{\"a}}}]{Massel2012}%
  \BibitemOpen
  \bibfield  {author} {\bibinfo {author} {\bibfnamefont {F.}~\bibnamefont
  {Massel}}, \bibinfo {author} {\bibfnamefont {S.~U.}\ \bibnamefont {Cho}},
  \bibinfo {author} {\bibfnamefont {J.-M.}\ \bibnamefont {Pirkkalainen}},
  \bibinfo {author} {\bibfnamefont {P.~J.}\ \bibnamefont {Hakonen}}, \bibinfo
  {author} {\bibfnamefont {T.~T.}\ \bibnamefont {Heikkil{\"a}}},\ and\ \bibinfo
  {author} {\bibfnamefont {M.~A.}\ \bibnamefont {Sillanp{\"a}{\"a}}},\
  }\bibfield  {title} {\bibinfo {title} {Multimode circuit optomechanics near
  the quantum limit},\ }\href {https://doi.org/10.1038/ncomms1993} {\bibfield
  {journal} {\bibinfo  {journal} {Nat. Commun.}\ }\textbf {\bibinfo {volume}
  {3}},\ \bibinfo {pages} {987} (\bibinfo {year} {2012})}\BibitemShut {NoStop}%
\bibitem [{\citenamefont {McClelland}\ \emph {et~al.}(2011)\citenamefont
  {McClelland}, \citenamefont {Mavalvala}, \citenamefont {Chen},\ and\
  \citenamefont {Schnabel}}]{LPR2011}%
  \BibitemOpen
  \bibfield  {author} {\bibinfo {author} {\bibfnamefont {D.~E.}\ \bibnamefont
  {McClelland}}, \bibinfo {author} {\bibfnamefont {N.}~\bibnamefont
  {Mavalvala}}, \bibinfo {author} {\bibfnamefont {Y.}~\bibnamefont {Chen}},\
  and\ \bibinfo {author} {\bibfnamefont {R.}~\bibnamefont {Schnabel}},\
  }\bibfield  {title} {\bibinfo {title} {Advanced interferometry, quantum
  optics and optomechanics in gravitational wave detectors},\ }\href
  {https://doi.org/10.1002/lpor.201000034} {\bibfield  {journal} {\bibinfo
  {journal} {Laser Photon. Rev.}\ }\textbf {\bibinfo {volume} {5}},\ \bibinfo
  {pages} {677} (\bibinfo {year} {2011})}\BibitemShut {NoStop}%
\bibitem [{\citenamefont {Li}\ \emph {et~al.}(2022)\citenamefont {Li},
  \citenamefont {Lu}, \citenamefont {Wang}, \citenamefont {Xin},\ and\
  \citenamefont {Li}}]{Li:22}%
  \BibitemOpen
  \bibfield  {author} {\bibinfo {author} {\bibfnamefont {G.}~\bibnamefont
  {Li}}, \bibinfo {author} {\bibfnamefont {X.-M.}\ \bibnamefont {Lu}}, \bibinfo
  {author} {\bibfnamefont {X.}~\bibnamefont {Wang}}, \bibinfo {author}
  {\bibfnamefont {J.}~\bibnamefont {Xin}},\ and\ \bibinfo {author}
  {\bibfnamefont {X.}~\bibnamefont {Li}},\ }\bibfield  {title} {\bibinfo
  {title} {Optomechanical gyroscope simultaneously estimating the position of
  the rotation axis},\ }\href {https://doi.org/10.1364/JOSAB.441232} {\bibfield
   {journal} {\bibinfo  {journal} {J. Opt. Soc. Am. B}\ }\textbf {\bibinfo
  {volume} {39}},\ \bibinfo {pages} {98} (\bibinfo {year} {2022})}\BibitemShut
  {NoStop}%
\bibitem [{\citenamefont {Sun}\ \emph {et~al.}(2024)\citenamefont {Sun},
  \citenamefont {Kovanis}, \citenamefont {Lončar},\ and\ \citenamefont
  {Lin}}]{sun2023}%
  \BibitemOpen
  \bibfield  {author} {\bibinfo {author} {\bibfnamefont {M.}~\bibnamefont
  {Sun}}, \bibinfo {author} {\bibfnamefont {V.}~\bibnamefont {Kovanis}},
  \bibinfo {author} {\bibfnamefont {M.}~\bibnamefont {Lončar}},\ and\ \bibinfo
  {author} {\bibfnamefont {Z.}~\bibnamefont {Lin}},\ }\bibfield  {title}
  {\bibinfo {title} {Bayesian optimization of fisher information in nonlinear
  multiresonant quantum photonics gyroscopes},\ }\href
  {https://doi.org/doi:10.1515/nanoph-2024-0032} {\bibfield  {journal}
  {\bibinfo  {journal} {Nanophotonics}\ }\textbf {\bibinfo {volume} {13}},\
  \bibinfo {pages} {2401} (\bibinfo {year} {2024})}\BibitemShut {NoStop}%
\bibitem [{\citenamefont {Maayani}\ \emph {et~al.}(2018)\citenamefont
  {Maayani}, \citenamefont {Dahan}, \citenamefont {Kligerman}, \citenamefont
  {Moses}, \citenamefont {Hassan}, \citenamefont {Jing}, \citenamefont {Nori},
  \citenamefont {Christodoulides},\ and\ \citenamefont {Carmon}}]{Maayani2018}%
  \BibitemOpen
  \bibfield  {author} {\bibinfo {author} {\bibfnamefont {S.}~\bibnamefont
  {Maayani}}, \bibinfo {author} {\bibfnamefont {R.}~\bibnamefont {Dahan}},
  \bibinfo {author} {\bibfnamefont {Y.}~\bibnamefont {Kligerman}}, \bibinfo
  {author} {\bibfnamefont {E.}~\bibnamefont {Moses}}, \bibinfo {author}
  {\bibfnamefont {A.~U.}\ \bibnamefont {Hassan}}, \bibinfo {author}
  {\bibfnamefont {H.}~\bibnamefont {Jing}}, \bibinfo {author} {\bibfnamefont
  {F.}~\bibnamefont {Nori}}, \bibinfo {author} {\bibfnamefont {D.~N.}\
  \bibnamefont {Christodoulides}},\ and\ \bibinfo {author} {\bibfnamefont
  {T.}~\bibnamefont {Carmon}},\ }\bibfield  {title} {\bibinfo {title} {Flying
  couplers above spinning resonators generate irreversible refraction},\ }\href
  {https://doi.org/10.1038/s41586-018-0245-5} {\bibfield  {journal} {\bibinfo
  {journal} {Nature (London)}\ }\textbf {\bibinfo {volume} {558}},\ \bibinfo
  {pages} {569} (\bibinfo {year} {2018})}\BibitemShut {NoStop}%
\bibitem [{\citenamefont {Jiao}\ \emph {et~al.}(2020)\citenamefont {Jiao},
  \citenamefont {Zhang}, \citenamefont {Zhang}, \citenamefont {Miranowicz},
  \citenamefont {Kuang},\ and\ \citenamefont {Jing}}]{PhysRevLett.125.143605}%
  \BibitemOpen
  \bibfield  {author} {\bibinfo {author} {\bibfnamefont {Y.-F.}\ \bibnamefont
  {Jiao}}, \bibinfo {author} {\bibfnamefont {S.-D.}\ \bibnamefont {Zhang}},
  \bibinfo {author} {\bibfnamefont {Y.-L.}\ \bibnamefont {Zhang}}, \bibinfo
  {author} {\bibfnamefont {A.}~\bibnamefont {Miranowicz}}, \bibinfo {author}
  {\bibfnamefont {L.-M.}\ \bibnamefont {Kuang}},\ and\ \bibinfo {author}
  {\bibfnamefont {H.}~\bibnamefont {Jing}},\ }\bibfield  {title} {\bibinfo
  {title} {Nonreciprocal optomechanical entanglement against backscattering
  losses},\ }\href {https://doi.org/10.1103/PhysRevLett.125.143605} {\bibfield
  {journal} {\bibinfo  {journal} {Phys. Rev. Lett.}\ }\textbf {\bibinfo
  {volume} {125}},\ \bibinfo {pages} {143605} (\bibinfo {year}
  {2020})}\BibitemShut {NoStop}%
\bibitem [{\citenamefont {Jiao}\ \emph {et~al.}(2022)\citenamefont {Jiao},
  \citenamefont {Liu}, \citenamefont {Li}, \citenamefont {Yang}, \citenamefont
  {Kuang},\ and\ \citenamefont {Jing}}]{PhysRevApplied.18.064008}%
  \BibitemOpen
  \bibfield  {author} {\bibinfo {author} {\bibfnamefont {Y.-F.}\ \bibnamefont
  {Jiao}}, \bibinfo {author} {\bibfnamefont {J.-X.}\ \bibnamefont {Liu}},
  \bibinfo {author} {\bibfnamefont {Y.}~\bibnamefont {Li}}, \bibinfo {author}
  {\bibfnamefont {R.}~\bibnamefont {Yang}}, \bibinfo {author} {\bibfnamefont
  {L.-M.}\ \bibnamefont {Kuang}},\ and\ \bibinfo {author} {\bibfnamefont
  {H.}~\bibnamefont {Jing}},\ }\bibfield  {title} {\bibinfo {title}
  {Nonreciprocal enhancement of remote entanglement between nonidentical
  mechanical oscillators},\ }\href
  {https://doi.org/10.1103/PhysRevApplied.18.064008} {\bibfield  {journal}
  {\bibinfo  {journal} {Phys. Rev. Appl.}\ }\textbf {\bibinfo {volume} {18}},\
  \bibinfo {pages} {064008} (\bibinfo {year} {2022})}\BibitemShut {NoStop}%
\bibitem [{\citenamefont {Huang}\ \emph {et~al.}(2018)\citenamefont {Huang},
  \citenamefont {Miranowicz}, \citenamefont {Liao}, \citenamefont {Nori},\ and\
  \citenamefont {Jing}}]{PhysRevLett.121.153601}%
  \BibitemOpen
  \bibfield  {author} {\bibinfo {author} {\bibfnamefont {R.}~\bibnamefont
  {Huang}}, \bibinfo {author} {\bibfnamefont {A.}~\bibnamefont {Miranowicz}},
  \bibinfo {author} {\bibfnamefont {J.-Q.}\ \bibnamefont {Liao}}, \bibinfo
  {author} {\bibfnamefont {F.}~\bibnamefont {Nori}},\ and\ \bibinfo {author}
  {\bibfnamefont {H.}~\bibnamefont {Jing}},\ }\bibfield  {title} {\bibinfo
  {title} {Nonreciprocal photon blockade},\ }\href
  {https://doi.org/10.1103/PhysRevLett.121.153601} {\bibfield  {journal}
  {\bibinfo  {journal} {Phys. Rev. Lett.}\ }\textbf {\bibinfo {volume} {121}},\
  \bibinfo {pages} {153601} (\bibinfo {year} {2018})}\BibitemShut {NoStop}%
\bibitem [{\citenamefont {Zhu}\ \emph {et~al.}(2024)\citenamefont {Zhu},
  \citenamefont {Hu}, \citenamefont {Wang}, \citenamefont {Qin}, \citenamefont
  {L\"u},\ and\ \citenamefont {Nori}}]{PhysRevLett.132.193602}%
  \BibitemOpen
  \bibfield  {author} {\bibinfo {author} {\bibfnamefont {G.-L.}\ \bibnamefont
  {Zhu}}, \bibinfo {author} {\bibfnamefont {C.-S.}\ \bibnamefont {Hu}},
  \bibinfo {author} {\bibfnamefont {H.}~\bibnamefont {Wang}}, \bibinfo {author}
  {\bibfnamefont {W.}~\bibnamefont {Qin}}, \bibinfo {author} {\bibfnamefont
  {X.-Y.}\ \bibnamefont {L\"u}},\ and\ \bibinfo {author} {\bibfnamefont
  {F.}~\bibnamefont {Nori}},\ }\bibfield  {title} {\bibinfo {title}
  {Nonreciprocal superradiant phase transitions and multicriticality in a
  cavity qed system},\ }\href {https://doi.org/10.1103/PhysRevLett.132.193602}
  {\bibfield  {journal} {\bibinfo  {journal} {Phys. Rev. Lett.}\ }\textbf
  {\bibinfo {volume} {132}},\ \bibinfo {pages} {193602} (\bibinfo {year}
  {2024})}\BibitemShut {NoStop}%
\bibitem [{\citenamefont {Bin}\ \emph {et~al.}(2024)\citenamefont {Bin},
  \citenamefont {Jing}, \citenamefont {Wu}, \citenamefont {Nori},\ and\
  \citenamefont {L\"u}}]{PhysRevLett.133.043601}%
  \BibitemOpen
  \bibfield  {author} {\bibinfo {author} {\bibfnamefont {Q.}~\bibnamefont
  {Bin}}, \bibinfo {author} {\bibfnamefont {H.}~\bibnamefont {Jing}}, \bibinfo
  {author} {\bibfnamefont {Y.}~\bibnamefont {Wu}}, \bibinfo {author}
  {\bibfnamefont {F.}~\bibnamefont {Nori}},\ and\ \bibinfo {author}
  {\bibfnamefont {X.-Y.}\ \bibnamefont {L\"u}},\ }\bibfield  {title} {\bibinfo
  {title} {Nonreciprocal bundle emissions of quantum entangled pairs},\ }\href
  {https://doi.org/10.1103/PhysRevLett.133.043601} {\bibfield  {journal}
  {\bibinfo  {journal} {Phys. Rev. Lett.}\ }\textbf {\bibinfo {volume} {133}},\
  \bibinfo {pages} {043601} (\bibinfo {year} {2024})}\BibitemShut {NoStop}%
\bibitem [{\citenamefont {Jing}\ \emph {et~al.}(2018)\citenamefont {Jing},
  \citenamefont {L\"{u}}, \citenamefont {\"{O}zdemir}, \citenamefont {Carmon},\
  and\ \citenamefont {Nori}}]{Jing:18}%
  \BibitemOpen
  \bibfield  {author} {\bibinfo {author} {\bibfnamefont {H.}~\bibnamefont
  {Jing}}, \bibinfo {author} {\bibfnamefont {H.}~\bibnamefont {L\"{u}}},
  \bibinfo {author} {\bibfnamefont {S.~K.}\ \bibnamefont {\"{O}zdemir}},
  \bibinfo {author} {\bibfnamefont {T.}~\bibnamefont {Carmon}},\ and\ \bibinfo
  {author} {\bibfnamefont {F.}~\bibnamefont {Nori}},\ }\bibfield  {title}
  {\bibinfo {title} {Nanoparticle sensing with a spinning resonator},\ }\href
  {https://doi.org/10.1364/OPTICA.5.001424} {\bibfield  {journal} {\bibinfo
  {journal} {Optica}\ }\textbf {\bibinfo {volume} {5}},\ \bibinfo {pages}
  {1424} (\bibinfo {year} {2018})}\BibitemShut {NoStop}%
\bibitem [{\citenamefont {Zhang}\ \emph {et~al.}(2020)\citenamefont {Zhang},
  \citenamefont {Huang}, \citenamefont {Zhang}, \citenamefont {Li},
  \citenamefont {Qiu}, \citenamefont {Nori},\ and\ \citenamefont
  {Jing}}]{Zhang2020}%
  \BibitemOpen
  \bibfield  {author} {\bibinfo {author} {\bibfnamefont {H.}~\bibnamefont
  {Zhang}}, \bibinfo {author} {\bibfnamefont {R.}~\bibnamefont {Huang}},
  \bibinfo {author} {\bibfnamefont {S.-D.}\ \bibnamefont {Zhang}}, \bibinfo
  {author} {\bibfnamefont {Y.}~\bibnamefont {Li}}, \bibinfo {author}
  {\bibfnamefont {C.-W.}\ \bibnamefont {Qiu}}, \bibinfo {author} {\bibfnamefont
  {F.}~\bibnamefont {Nori}},\ and\ \bibinfo {author} {\bibfnamefont
  {H.}~\bibnamefont {Jing}},\ }\bibfield  {title} {\bibinfo {title} {Breaking
  anti-pt symmetry by spinning a resonator},\ }\href
  {https://doi.org/10.1021/acs.nanolett.0c03119} {\bibfield  {journal}
  {\bibinfo  {journal} {Nano Letters}\ }\textbf {\bibinfo {volume} {20}},\
  \bibinfo {pages} {7594} (\bibinfo {year} {2020})}\BibitemShut {NoStop}%
\bibitem [{\citenamefont {Wang}\ \emph {et~al.}(2024)\citenamefont {Wang},
  \citenamefont {Zhang}, \citenamefont {Jiao}, \citenamefont {Zhang},
  \citenamefont {Lu}, \citenamefont {Li}, \citenamefont {Qiu},\ and\
  \citenamefont {Jing}}]{Appl.Phys.Rev.11.031409}%
  \BibitemOpen
  \bibfield  {author} {\bibinfo {author} {\bibfnamefont {J.}~\bibnamefont
  {Wang}}, \bibinfo {author} {\bibfnamefont {Q.}~\bibnamefont {Zhang}},
  \bibinfo {author} {\bibfnamefont {Y.-F.}\ \bibnamefont {Jiao}}, \bibinfo
  {author} {\bibfnamefont {S.-D.}\ \bibnamefont {Zhang}}, \bibinfo {author}
  {\bibfnamefont {T.-X.}\ \bibnamefont {Lu}}, \bibinfo {author} {\bibfnamefont
  {Z.}~\bibnamefont {Li}}, \bibinfo {author} {\bibfnamefont {C.-W.}\
  \bibnamefont {Qiu}},\ and\ \bibinfo {author} {\bibfnamefont {H.}~\bibnamefont
  {Jing}},\ }\bibfield  {title} {\bibinfo {title} {Quantum advantage of one-way
  squeezing in weak-force sensing},\ }\href {https://doi.org/10.1063/5.0208107}
  {\bibfield  {journal} {\bibinfo  {journal} {Appl. Phys. Rev.}\ }\textbf
  {\bibinfo {volume} {11}},\ \bibinfo {pages} {031409} (\bibinfo {year}
  {2024})}\BibitemShut {NoStop}%
\bibitem [{\citenamefont {Helstrom}(1969)}]{Helstrom1969}%
  \BibitemOpen
  \bibfield  {author} {\bibinfo {author} {\bibfnamefont {C.~W.}\ \bibnamefont
  {Helstrom}},\ }\bibfield  {title} {\bibinfo {title} {Quantum detection and
  estimation theory},\ }\href {https://doi.org/10.1007/BF01007479} {\bibfield
  {journal} {\bibinfo  {journal} {J. Stat. Phys.}\ }\textbf {\bibinfo {volume}
  {1}},\ \bibinfo {pages} {231} (\bibinfo {year} {1969})}\BibitemShut {NoStop}%
\bibitem [{\citenamefont {Xu}\ \emph {et~al.}(2019)\citenamefont {Xu},
  \citenamefont {Li}, \citenamefont {Liu}, \citenamefont {Wang}, \citenamefont
  {Yuan},\ and\ \citenamefont {Wang}}]{Xu2019}%
  \BibitemOpen
  \bibfield  {author} {\bibinfo {author} {\bibfnamefont {H.}~\bibnamefont
  {Xu}}, \bibinfo {author} {\bibfnamefont {J.}~\bibnamefont {Li}}, \bibinfo
  {author} {\bibfnamefont {L.}~\bibnamefont {Liu}}, \bibinfo {author}
  {\bibfnamefont {Y.}~\bibnamefont {Wang}}, \bibinfo {author} {\bibfnamefont
  {H.}~\bibnamefont {Yuan}},\ and\ \bibinfo {author} {\bibfnamefont
  {X.}~\bibnamefont {Wang}},\ }\bibfield  {title} {\bibinfo {title}
  {Generalizable control for quantum parameter estimation through reinforcement
  learning},\ }\href {https://doi.org/10.1038/s41534-019-0198-z} {\bibfield
  {journal} {\bibinfo  {journal} {npj Quantum Inf.}\ }\textbf {\bibinfo
  {volume} {5}},\ \bibinfo {pages} {82} (\bibinfo {year} {2019})}\BibitemShut
  {NoStop}%
\bibitem [{\citenamefont {Wang}\ \emph {et~al.}(2020)\citenamefont {Wang},
  \citenamefont {Ashida},\ and\ \citenamefont {Ueda}}]{PhysRevLett.125.100401}%
  \BibitemOpen
  \bibfield  {author} {\bibinfo {author} {\bibfnamefont {Z.~T.}\ \bibnamefont
  {Wang}}, \bibinfo {author} {\bibfnamefont {Y.}~\bibnamefont {Ashida}},\ and\
  \bibinfo {author} {\bibfnamefont {M.}~\bibnamefont {Ueda}},\ }\bibfield
  {title} {\bibinfo {title} {Deep reinforcement learning control of quantum
  cartpoles},\ }\href {https://doi.org/10.1103/PhysRevLett.125.100401}
  {\bibfield  {journal} {\bibinfo  {journal} {Phys. Rev. Lett.}\ }\textbf
  {\bibinfo {volume} {125}},\ \bibinfo {pages} {100401} (\bibinfo {year}
  {2020})}\BibitemShut {NoStop}%
\bibitem [{\citenamefont {Porotti}\ \emph {et~al.}(2019)\citenamefont
  {Porotti}, \citenamefont {Tamascelli}, \citenamefont {Restelli},\ and\
  \citenamefont {Prati}}]{Porotti2019}%
  \BibitemOpen
  \bibfield  {author} {\bibinfo {author} {\bibfnamefont {R.}~\bibnamefont
  {Porotti}}, \bibinfo {author} {\bibfnamefont {D.}~\bibnamefont {Tamascelli}},
  \bibinfo {author} {\bibfnamefont {M.}~\bibnamefont {Restelli}},\ and\
  \bibinfo {author} {\bibfnamefont {E.}~\bibnamefont {Prati}},\ }\bibfield
  {title} {\bibinfo {title} {Coherent transport of quantum states by deep
  reinforcement learning},\ }\href {https://doi.org/10.1038/s42005-019-0169-x}
  {\bibfield  {journal} {\bibinfo  {journal} {Commun. Phys.}\ }\textbf
  {\bibinfo {volume} {2}},\ \bibinfo {pages} {61} (\bibinfo {year}
  {2019})}\BibitemShut {NoStop}%
\bibitem [{\citenamefont {Guo}\ \emph {et~al.}(2021)\citenamefont {Guo},
  \citenamefont {Chen}, \citenamefont {Liu}, \citenamefont {Xue}, \citenamefont
  {Chen}, \citenamefont {Cao}, \citenamefont {Mao}, \citenamefont {Tey},\ and\
  \citenamefont {You}}]{PhysRevLett.126.060401}%
  \BibitemOpen
  \bibfield  {author} {\bibinfo {author} {\bibfnamefont {S.-F.}\ \bibnamefont
  {Guo}}, \bibinfo {author} {\bibfnamefont {F.}~\bibnamefont {Chen}}, \bibinfo
  {author} {\bibfnamefont {Q.}~\bibnamefont {Liu}}, \bibinfo {author}
  {\bibfnamefont {M.}~\bibnamefont {Xue}}, \bibinfo {author} {\bibfnamefont
  {J.-J.}\ \bibnamefont {Chen}}, \bibinfo {author} {\bibfnamefont {J.-H.}\
  \bibnamefont {Cao}}, \bibinfo {author} {\bibfnamefont {T.-W.}\ \bibnamefont
  {Mao}}, \bibinfo {author} {\bibfnamefont {M.~K.}\ \bibnamefont {Tey}},\ and\
  \bibinfo {author} {\bibfnamefont {L.}~\bibnamefont {You}},\ }\bibfield
  {title} {\bibinfo {title} {Faster state preparation across quantum phase
  transition assisted by reinforcement learning},\ }\href
  {https://doi.org/10.1103/PhysRevLett.126.060401} {\bibfield  {journal}
  {\bibinfo  {journal} {Phys. Rev. Lett.}\ }\textbf {\bibinfo {volume} {126}},\
  \bibinfo {pages} {060401} (\bibinfo {year} {2021})}\BibitemShut {NoStop}%
\bibitem [{\citenamefont {F\"osel}\ \emph {et~al.}(2018)\citenamefont
  {F\"osel}, \citenamefont {Tighineanu}, \citenamefont {Weiss},\ and\
  \citenamefont {Marquardt}}]{PhysRevX.8.031084}%
  \BibitemOpen
  \bibfield  {author} {\bibinfo {author} {\bibfnamefont {T.}~\bibnamefont
  {F\"osel}}, \bibinfo {author} {\bibfnamefont {P.}~\bibnamefont {Tighineanu}},
  \bibinfo {author} {\bibfnamefont {T.}~\bibnamefont {Weiss}},\ and\ \bibinfo
  {author} {\bibfnamefont {F.}~\bibnamefont {Marquardt}},\ }\bibfield  {title}
  {\bibinfo {title} {Reinforcement learning with neural networks for quantum
  feedback},\ }\href {https://doi.org/10.1103/PhysRevX.8.031084} {\bibfield
  {journal} {\bibinfo  {journal} {Phys. Rev. X}\ }\textbf {\bibinfo {volume}
  {8}},\ \bibinfo {pages} {031084} (\bibinfo {year} {2018})}\BibitemShut
  {NoStop}%
\bibitem [{\citenamefont {Ai}\ \emph {et~al.}(2022)\citenamefont {Ai},
  \citenamefont {Ding}, \citenamefont {Ban}, \citenamefont
  {Mart{\'i}n-Guerrero}, \citenamefont {Casanova}, \citenamefont {Cui},
  \citenamefont {Huang}, \citenamefont {Chen}, \citenamefont {Li},\ and\
  \citenamefont {Guo}}]{Ai2022}%
  \BibitemOpen
  \bibfield  {author} {\bibinfo {author} {\bibfnamefont {M.-Z.}\ \bibnamefont
  {Ai}}, \bibinfo {author} {\bibfnamefont {Y.}~\bibnamefont {Ding}}, \bibinfo
  {author} {\bibfnamefont {Y.}~\bibnamefont {Ban}}, \bibinfo {author}
  {\bibfnamefont {J.~D.}\ \bibnamefont {Mart{\'i}n-Guerrero}}, \bibinfo
  {author} {\bibfnamefont {J.}~\bibnamefont {Casanova}}, \bibinfo {author}
  {\bibfnamefont {J.-M.}\ \bibnamefont {Cui}}, \bibinfo {author} {\bibfnamefont
  {Y.-F.}\ \bibnamefont {Huang}}, \bibinfo {author} {\bibfnamefont
  {X.}~\bibnamefont {Chen}}, \bibinfo {author} {\bibfnamefont {C.-F.}\
  \bibnamefont {Li}},\ and\ \bibinfo {author} {\bibfnamefont {G.-C.}\
  \bibnamefont {Guo}},\ }\bibfield  {title} {\bibinfo {title} {Experimentally
  realizing efficient quantum control with reinforcement learning},\ }\href
  {https://doi.org/10.1007/s11433-021-1841-2} {\bibfield  {journal} {\bibinfo
  {journal} {Sci. China Phys. Mech. Astron.}\ }\textbf {\bibinfo {volume}
  {65}},\ \bibinfo {pages} {250312} (\bibinfo {year} {2022})}\BibitemShut
  {NoStop}%
\bibitem [{\citenamefont {Zhang}\ \emph {et~al.}(2023)\citenamefont {Zhang},
  \citenamefont {Li}, \citenamefont {Tan}, \citenamefont {Bu}, \citenamefont
  {Yuan}, \citenamefont {Wang}, \citenamefont {Ding}, \citenamefont {Ding},
  \citenamefont {Chen}, \citenamefont {Yan}, \citenamefont {Su}, \citenamefont
  {Xiong}, \citenamefont {Zhou},\ and\ \citenamefont {Feng}}]{Zhang2023}%
  \BibitemOpen
  \bibfield  {author} {\bibinfo {author} {\bibfnamefont {J.}~\bibnamefont
  {Zhang}}, \bibinfo {author} {\bibfnamefont {J.}~\bibnamefont {Li}}, \bibinfo
  {author} {\bibfnamefont {Q.-S.}\ \bibnamefont {Tan}}, \bibinfo {author}
  {\bibfnamefont {J.}~\bibnamefont {Bu}}, \bibinfo {author} {\bibfnamefont
  {W.}~\bibnamefont {Yuan}}, \bibinfo {author} {\bibfnamefont {B.}~\bibnamefont
  {Wang}}, \bibinfo {author} {\bibfnamefont {G.}~\bibnamefont {Ding}}, \bibinfo
  {author} {\bibfnamefont {W.}~\bibnamefont {Ding}}, \bibinfo {author}
  {\bibfnamefont {L.}~\bibnamefont {Chen}}, \bibinfo {author} {\bibfnamefont
  {L.}~\bibnamefont {Yan}}, \bibinfo {author} {\bibfnamefont {S.}~\bibnamefont
  {Su}}, \bibinfo {author} {\bibfnamefont {T.}~\bibnamefont {Xiong}}, \bibinfo
  {author} {\bibfnamefont {F.}~\bibnamefont {Zhou}},\ and\ \bibinfo {author}
  {\bibfnamefont {M.}~\bibnamefont {Feng}},\ }\bibfield  {title} {\bibinfo
  {title} {Single-atom exploration of optimized nonequilibrium quantum
  thermodynamics by reinforcement learning},\ }\href
  {https://doi.org/10.1038/s42005-023-01408-5} {\bibfield  {journal} {\bibinfo
  {journal} {Commun. Phys.}\ }\textbf {\bibinfo {volume} {6}},\ \bibinfo
  {pages} {286} (\bibinfo {year} {2023})}\BibitemShut {NoStop}%
\bibitem [{\citenamefont {Zhang}\ \emph {et~al.}(2024)\citenamefont {Zhang},
  \citenamefont {Bu}, \citenamefont {Li}, \citenamefont {Meng}, \citenamefont
  {Ding}, \citenamefont {Wang}, \citenamefont {Yuan}, \citenamefont {Du},
  \citenamefont {Ding}, \citenamefont {Chen}, \citenamefont {Chen},
  \citenamefont {Zhou}, \citenamefont {Xu},\ and\ \citenamefont
  {Feng}}]{PhysRevLett.132.213602}%
  \BibitemOpen
  \bibfield  {author} {\bibinfo {author} {\bibfnamefont {J.-W.}\ \bibnamefont
  {Zhang}}, \bibinfo {author} {\bibfnamefont {J.-T.}\ \bibnamefont {Bu}},
  \bibinfo {author} {\bibfnamefont {J.~C.}\ \bibnamefont {Li}}, \bibinfo
  {author} {\bibfnamefont {W.}~\bibnamefont {Meng}}, \bibinfo {author}
  {\bibfnamefont {W.-Q.}\ \bibnamefont {Ding}}, \bibinfo {author}
  {\bibfnamefont {B.}~\bibnamefont {Wang}}, \bibinfo {author} {\bibfnamefont
  {W.-F.}\ \bibnamefont {Yuan}}, \bibinfo {author} {\bibfnamefont {H.-J.}\
  \bibnamefont {Du}}, \bibinfo {author} {\bibfnamefont {G.-Y.}\ \bibnamefont
  {Ding}}, \bibinfo {author} {\bibfnamefont {W.-J.}\ \bibnamefont {Chen}},
  \bibinfo {author} {\bibfnamefont {L.}~\bibnamefont {Chen}}, \bibinfo {author}
  {\bibfnamefont {F.}~\bibnamefont {Zhou}}, \bibinfo {author} {\bibfnamefont
  {Z.}~\bibnamefont {Xu}},\ and\ \bibinfo {author} {\bibfnamefont
  {M.}~\bibnamefont {Feng}},\ }\bibfield  {title} {\bibinfo {title}
  {Single-atom verification of the optimal trade-off between speed and cost in
  shortcuts to adiabaticity},\ }\href
  {https://doi.org/10.1103/PhysRevLett.132.213602} {\bibfield  {journal}
  {\bibinfo  {journal} {Phys. Rev. Lett.}\ }\textbf {\bibinfo {volume} {132}},\
  \bibinfo {pages} {213602} (\bibinfo {year} {2024})}\BibitemShut {NoStop}%
\bibitem [{\citenamefont {Tan}\ \emph {et~al.}(2021)\citenamefont {Tan},
  \citenamefont {Zhang}, \citenamefont {Chen}, \citenamefont {Liao},\ and\
  \citenamefont {Liu}}]{PhysRevA.103.032601}%
  \BibitemOpen
  \bibfield  {author} {\bibinfo {author} {\bibfnamefont {Q.-S.}\ \bibnamefont
  {Tan}}, \bibinfo {author} {\bibfnamefont {M.}~\bibnamefont {Zhang}}, \bibinfo
  {author} {\bibfnamefont {Y.}~\bibnamefont {Chen}}, \bibinfo {author}
  {\bibfnamefont {J.-Q.}\ \bibnamefont {Liao}},\ and\ \bibinfo {author}
  {\bibfnamefont {J.}~\bibnamefont {Liu}},\ }\bibfield  {title} {\bibinfo
  {title} {Generation and storage of spin squeezing via learning-assisted
  optimal control},\ }\href {https://doi.org/10.1103/PhysRevA.103.032601}
  {\bibfield  {journal} {\bibinfo  {journal} {Phys. Rev. A}\ }\textbf {\bibinfo
  {volume} {103}},\ \bibinfo {pages} {032601} (\bibinfo {year}
  {2021})}\BibitemShut {NoStop}%
\bibitem [{\citenamefont {Bukov}\ \emph {et~al.}(2018)\citenamefont {Bukov},
  \citenamefont {Day}, \citenamefont {Sels}, \citenamefont {Weinberg},
  \citenamefont {Polkovnikov},\ and\ \citenamefont
  {Mehta}}]{PhysRevX.8.031086}%
  \BibitemOpen
  \bibfield  {author} {\bibinfo {author} {\bibfnamefont {M.}~\bibnamefont
  {Bukov}}, \bibinfo {author} {\bibfnamefont {A.~G.~R.}\ \bibnamefont {Day}},
  \bibinfo {author} {\bibfnamefont {D.}~\bibnamefont {Sels}}, \bibinfo {author}
  {\bibfnamefont {P.}~\bibnamefont {Weinberg}}, \bibinfo {author}
  {\bibfnamefont {A.}~\bibnamefont {Polkovnikov}},\ and\ \bibinfo {author}
  {\bibfnamefont {P.}~\bibnamefont {Mehta}},\ }\bibfield  {title} {\bibinfo
  {title} {Reinforcement learning in different phases of quantum control},\
  }\href {https://doi.org/10.1103/PhysRevX.8.031086} {\bibfield  {journal}
  {\bibinfo  {journal} {Phys. Rev. X}\ }\textbf {\bibinfo {volume} {8}},\
  \bibinfo {pages} {031086} (\bibinfo {year} {2018})}\BibitemShut {NoStop}%
\bibitem [{\citenamefont {Zeng}\ \emph {et~al.}(2023)\citenamefont {Zeng},
  \citenamefont {Zhou}, \citenamefont {Rinaldi}, \citenamefont {Gneiting},\
  and\ \citenamefont {Nori}}]{PhysRevLett.131.050601}%
  \BibitemOpen
  \bibfield  {author} {\bibinfo {author} {\bibfnamefont {Y.}~\bibnamefont
  {Zeng}}, \bibinfo {author} {\bibfnamefont {Z.-Y.}\ \bibnamefont {Zhou}},
  \bibinfo {author} {\bibfnamefont {E.}~\bibnamefont {Rinaldi}}, \bibinfo
  {author} {\bibfnamefont {C.}~\bibnamefont {Gneiting}},\ and\ \bibinfo
  {author} {\bibfnamefont {F.}~\bibnamefont {Nori}},\ }\bibfield  {title}
  {\bibinfo {title} {Approximate autonomous quantum error correction with
  reinforcement learning},\ }\href
  {https://doi.org/10.1103/PhysRevLett.131.050601} {\bibfield  {journal}
  {\bibinfo  {journal} {Phys. Rev. Lett.}\ }\textbf {\bibinfo {volume} {131}},\
  \bibinfo {pages} {050601} (\bibinfo {year} {2023})}\BibitemShut {NoStop}%
\bibitem [{\citenamefont {Xiao}\ \emph {et~al.}(2022)\citenamefont {Xiao},
  \citenamefont {Fan},\ and\ \citenamefont {Zeng}}]{Xiao2022}%
  \BibitemOpen
  \bibfield  {author} {\bibinfo {author} {\bibfnamefont {T.}~\bibnamefont
  {Xiao}}, \bibinfo {author} {\bibfnamefont {J.}~\bibnamefont {Fan}},\ and\
  \bibinfo {author} {\bibfnamefont {G.}~\bibnamefont {Zeng}},\ }\bibfield
  {title} {\bibinfo {title} {Parameter estimation in quantum sensing based on
  deep reinforcement learning},\ }\href
  {https://doi.org/10.1038/s41534-021-00513-z} {\bibfield  {journal} {\bibinfo
  {journal} {npj Quantum Inf.}\ }\textbf {\bibinfo {volume} {8}},\ \bibinfo
  {pages} {2} (\bibinfo {year} {2022})}\BibitemShut {NoStop}%
\bibitem [{\citenamefont {Zhang}\ \emph {et~al.}(2022)\citenamefont {Zhang},
  \citenamefont {Yu}, \citenamefont {Yuan}, \citenamefont {Wang}, \citenamefont
  {Demkowicz-Dobrza\ifmmode~\acute{n}\else \'{n}\fi{}ski},\ and\ \citenamefont
  {Liu}}]{PhysRevResearch.4.043057}%
  \BibitemOpen
  \bibfield  {author} {\bibinfo {author} {\bibfnamefont {M.}~\bibnamefont
  {Zhang}}, \bibinfo {author} {\bibfnamefont {H.-M.}\ \bibnamefont {Yu}},
  \bibinfo {author} {\bibfnamefont {H.}~\bibnamefont {Yuan}}, \bibinfo {author}
  {\bibfnamefont {X.}~\bibnamefont {Wang}}, \bibinfo {author} {\bibfnamefont
  {R.}~\bibnamefont {Demkowicz-Dobrza\ifmmode~\acute{n}\else \'{n}\fi{}ski}},\
  and\ \bibinfo {author} {\bibfnamefont {J.}~\bibnamefont {Liu}},\ }\bibfield
  {title} {\bibinfo {title} {Quanestimation: An open-source toolkit for quantum
  parameter estimation},\ }\href
  {https://doi.org/10.1103/PhysRevResearch.4.043057} {\bibfield  {journal}
  {\bibinfo  {journal} {Phys. Rev. Res.}\ }\textbf {\bibinfo {volume} {4}},\
  \bibinfo {pages} {043057} (\bibinfo {year} {2022})}\BibitemShut {NoStop}%
\bibitem [{\citenamefont {Tan}\ \emph {et~al.}(2024)\citenamefont {Tan},
  \citenamefont {Liu}, \citenamefont {Xu}, \citenamefont {Wu},\ and\
  \citenamefont {Kuang}}]{PhysRevA.109.042417}%
  \BibitemOpen
  \bibfield  {author} {\bibinfo {author} {\bibfnamefont {Q.-S.}\ \bibnamefont
  {Tan}}, \bibinfo {author} {\bibfnamefont {X.}~\bibnamefont {Liu}}, \bibinfo
  {author} {\bibfnamefont {L.}~\bibnamefont {Xu}}, \bibinfo {author}
  {\bibfnamefont {W.}~\bibnamefont {Wu}},\ and\ \bibinfo {author}
  {\bibfnamefont {L.-M.}\ \bibnamefont {Kuang}},\ }\bibfield  {title} {\bibinfo
  {title} {Enhancement of sensitivity in low-temperature quantum thermometry
  via reinforcement learning},\ }\href
  {https://doi.org/10.1103/PhysRevA.109.042417} {\bibfield  {journal} {\bibinfo
   {journal} {Phys. Rev. A}\ }\textbf {\bibinfo {volume} {109}},\ \bibinfo
  {pages} {042417} (\bibinfo {year} {2024})}\BibitemShut {NoStop}%
\bibitem [{\citenamefont {Malykin}(2000)}]{Grigorii2000}%
  \BibitemOpen
  \bibfield  {author} {\bibinfo {author} {\bibfnamefont {G.~B.}\ \bibnamefont
  {Malykin}},\ }\bibfield  {title} {\bibinfo {title} {The sagnac effect:
  correct and incorrect explanations},\ }\href
  {https://doi.org/10.1070/PU2000v043n12ABEH000830} {\bibfield  {journal}
  {\bibinfo  {journal} {Physics-Uspekhi}\ }\textbf {\bibinfo {volume} {43}},\
  \bibinfo {pages} {1229} (\bibinfo {year} {2000})}\BibitemShut {NoStop}%
\bibitem [{\citenamefont {Breuer}\ and\ \citenamefont
  {Petruccione}(2002)}]{breuer2002theory}%
  \BibitemOpen
  \bibfield  {author} {\bibinfo {author} {\bibfnamefont {H.-P.}\ \bibnamefont
  {Breuer}}\ and\ \bibinfo {author} {\bibfnamefont {F.}~\bibnamefont
  {Petruccione}},\ }\href@noop {} {\emph {\bibinfo {title} {The theory of open
  quantum systems}}}\ (\bibinfo  {publisher} {OUP Oxford},\ \bibinfo {year}
  {2002})\BibitemShut {NoStop}%
\bibitem [{\citenamefont {Liu}\ \emph {et~al.}(2024)\citenamefont {Liu},
  \citenamefont {Zeng}, \citenamefont {Tan}, \citenamefont {Dong},
  \citenamefont {Nori},\ and\ \citenamefont {Liao}}]{PhysRevA.109.063508}%
  \BibitemOpen
  \bibfield  {author} {\bibinfo {author} {\bibfnamefont {Y.-H.}\ \bibnamefont
  {Liu}}, \bibinfo {author} {\bibfnamefont {Y.}~\bibnamefont {Zeng}}, \bibinfo
  {author} {\bibfnamefont {Q.-S.}\ \bibnamefont {Tan}}, \bibinfo {author}
  {\bibfnamefont {D.}~\bibnamefont {Dong}}, \bibinfo {author} {\bibfnamefont
  {F.}~\bibnamefont {Nori}},\ and\ \bibinfo {author} {\bibfnamefont {J.-Q.}\
  \bibnamefont {Liao}},\ }\bibfield  {title} {\bibinfo {title} {Optimal control
  of linear gaussian quantum systems via quantum learning control},\ }\href
  {https://doi.org/10.1103/PhysRevA.109.063508} {\bibfield  {journal} {\bibinfo
   {journal} {Phys. Rev. A}\ }\textbf {\bibinfo {volume} {109}},\ \bibinfo
  {pages} {063508} (\bibinfo {year} {2024})}\BibitemShut {NoStop}%
\bibitem [{\citenamefont {Liu}\ and\ \citenamefont
  {Liao}(2024)}]{PhysRevA.110.023519}%
  \BibitemOpen
  \bibfield  {author} {\bibinfo {author} {\bibfnamefont {Y.-H.}\ \bibnamefont
  {Liu}}\ and\ \bibinfo {author} {\bibfnamefont {J.-Q.}\ \bibnamefont {Liao}},\
  }\bibfield  {title} {\bibinfo {title} {Optimized mechanical quadrature
  squeezing beyond the 3-db limit via a gradient-descent algorithm},\ }\href
  {https://doi.org/10.1103/PhysRevA.110.023519} {\bibfield  {journal} {\bibinfo
   {journal} {Phys. Rev. A}\ }\textbf {\bibinfo {volume} {110}},\ \bibinfo
  {pages} {023519} (\bibinfo {year} {2024})}\BibitemShut {NoStop}%
\bibitem [{\citenamefont {Rao}(1992)}]{Rao1992}%
  \BibitemOpen
  \bibfield  {author} {\bibinfo {author} {\bibfnamefont {C.~R.}\ \bibnamefont
  {Rao}},\ }\bibinfo {title} {Information and the accuracy attainable in the
  estimation of statistical parameters},\ in\ \href
  {https://doi.org/10.1007/978-1-4612-0919-5_16} {\emph {\bibinfo {booktitle}
  {Breakthroughs in Statistics: Foundations and Basic Theory}}},\ \bibinfo
  {editor} {edited by\ \bibinfo {editor} {\bibfnamefont {S.}~\bibnamefont
  {Kotz}}\ and\ \bibinfo {editor} {\bibfnamefont {N.~L.}\ \bibnamefont
  {Johnson}}}\ (\bibinfo  {publisher} {Springer New York},\ \bibinfo {address}
  {New York, NY},\ \bibinfo {year} {1992})\ pp.\ \bibinfo {pages}
  {235--247}\BibitemShut {NoStop}%
\bibitem [{\citenamefont {\v{S}afr\'{a}nek}(2018)}]{Safranek_2019}%
  \BibitemOpen
  \bibfield  {author} {\bibinfo {author} {\bibfnamefont {D.}~\bibnamefont
  {\v{S}afr\'{a}nek}},\ }\bibfield  {title} {\bibinfo {title} {Estimation of
  gaussian quantum states},\ }\href {https://doi.org/10.1088/1751-8121/aaf068}
  {\bibfield  {journal} {\bibinfo  {journal} {J. Phys. A: Math. Theor.}\
  }\textbf {\bibinfo {volume} {52}},\ \bibinfo {pages} {035304} (\bibinfo
  {year} {2018})}\BibitemShut {NoStop}%
\bibitem [{\citenamefont {Liu}\ \emph {et~al.}(2019)\citenamefont {Liu},
  \citenamefont {Yuan}, \citenamefont {Lu},\ and\ \citenamefont
  {Wang}}]{Liu2020}%
  \BibitemOpen
  \bibfield  {author} {\bibinfo {author} {\bibfnamefont {J.}~\bibnamefont
  {Liu}}, \bibinfo {author} {\bibfnamefont {H.}~\bibnamefont {Yuan}}, \bibinfo
  {author} {\bibfnamefont {X.-M.}\ \bibnamefont {Lu}},\ and\ \bibinfo {author}
  {\bibfnamefont {X.}~\bibnamefont {Wang}},\ }\bibfield  {title} {\bibinfo
  {title} {Quantum fisher information matrix and multiparameter estimation},\
  }\href {https://doi.org/10.1088/1751-8121/ab5d4d} {\bibfield  {journal}
  {\bibinfo  {journal} {J. Phys. A: Math.Theor.}\ }\textbf {\bibinfo {volume}
  {53}},\ \bibinfo {pages} {023001} (\bibinfo {year} {2019})}\BibitemShut
  {NoStop}%
\bibitem [{\citenamefont {Riedinger}\ \emph {et~al.}(2018)\citenamefont
  {Riedinger}, \citenamefont {Wallucks}, \citenamefont {Marinkovi{\'{c}}},
  \citenamefont {L{\"o}schnauer}, \citenamefont {Aspelmeyer}, \citenamefont
  {Hong},\ and\ \citenamefont {Gr{\"o}blacher}}]{Riedinger2018}%
  \BibitemOpen
  \bibfield  {author} {\bibinfo {author} {\bibfnamefont {R.}~\bibnamefont
  {Riedinger}}, \bibinfo {author} {\bibfnamefont {A.}~\bibnamefont {Wallucks}},
  \bibinfo {author} {\bibfnamefont {I.}~\bibnamefont {Marinkovi{\'{c}}}},
  \bibinfo {author} {\bibfnamefont {C.}~\bibnamefont {L{\"o}schnauer}},
  \bibinfo {author} {\bibfnamefont {M.}~\bibnamefont {Aspelmeyer}}, \bibinfo
  {author} {\bibfnamefont {S.}~\bibnamefont {Hong}},\ and\ \bibinfo {author}
  {\bibfnamefont {S.}~\bibnamefont {Gr{\"o}blacher}},\ }\bibfield  {title}
  {\bibinfo {title} {Remote quantum entanglement between two micromechanical
  oscillators},\ }\href {https://doi.org/10.1038/s41586-018-0036-z} {\bibfield
  {journal} {\bibinfo  {journal} {Nature}\ }\textbf {\bibinfo {volume} {556}},\
  \bibinfo {pages} {473} (\bibinfo {year} {2018})}\BibitemShut {NoStop}%
\bibitem [{\citenamefont {Ockeloen-Korppi}\ \emph {et~al.}(2018)\citenamefont
  {Ockeloen-Korppi}, \citenamefont {Damsk{\"a}gg}, \citenamefont
  {Pirkkalainen}, \citenamefont {Asjad}, \citenamefont {Clerk}, \citenamefont
  {Massel}, \citenamefont {Woolley},\ and\ \citenamefont
  {Sillanp{\"a}{\"a}}}]{Ockeloen-Korppi2018}%
  \BibitemOpen
  \bibfield  {author} {\bibinfo {author} {\bibfnamefont {C.~F.}\ \bibnamefont
  {Ockeloen-Korppi}}, \bibinfo {author} {\bibfnamefont {E.}~\bibnamefont
  {Damsk{\"a}gg}}, \bibinfo {author} {\bibfnamefont {J.-M.}\ \bibnamefont
  {Pirkkalainen}}, \bibinfo {author} {\bibfnamefont {M.}~\bibnamefont {Asjad}},
  \bibinfo {author} {\bibfnamefont {A.~A.}\ \bibnamefont {Clerk}}, \bibinfo
  {author} {\bibfnamefont {F.}~\bibnamefont {Massel}}, \bibinfo {author}
  {\bibfnamefont {M.~J.}\ \bibnamefont {Woolley}},\ and\ \bibinfo {author}
  {\bibfnamefont {M.~A.}\ \bibnamefont {Sillanp{\"a}{\"a}}},\ }\bibfield
  {title} {\bibinfo {title} {Stabilized entanglement of massive mechanical
  oscillators},\ }\href {https://doi.org/10.1038/s41586-018-0038-x} {\bibfield
  {journal} {\bibinfo  {journal} {Nature}\ }\textbf {\bibinfo {volume} {556}},\
  \bibinfo {pages} {478} (\bibinfo {year} {2018})}\BibitemShut {NoStop}%
\bibitem [{\citenamefont {Kotler}\ \emph {et~al.}(2021)\citenamefont {Kotler},
  \citenamefont {Peterson}, \citenamefont {Shojaee}, \citenamefont {Lecocq},
  \citenamefont {Cicak}, \citenamefont {Kwiatkowski}, \citenamefont {Geller},
  \citenamefont {Glancy}, \citenamefont {Knill}, \citenamefont {Simmonds},
  \citenamefont {Aumentado},\ and\ \citenamefont
  {Teufel}}]{doi:10.1126/science.abf2998}%
  \BibitemOpen
  \bibfield  {author} {\bibinfo {author} {\bibfnamefont {S.}~\bibnamefont
  {Kotler}}, \bibinfo {author} {\bibfnamefont {G.~A.}\ \bibnamefont
  {Peterson}}, \bibinfo {author} {\bibfnamefont {E.}~\bibnamefont {Shojaee}},
  \bibinfo {author} {\bibfnamefont {F.}~\bibnamefont {Lecocq}}, \bibinfo
  {author} {\bibfnamefont {K.}~\bibnamefont {Cicak}}, \bibinfo {author}
  {\bibfnamefont {A.}~\bibnamefont {Kwiatkowski}}, \bibinfo {author}
  {\bibfnamefont {S.}~\bibnamefont {Geller}}, \bibinfo {author} {\bibfnamefont
  {S.}~\bibnamefont {Glancy}}, \bibinfo {author} {\bibfnamefont
  {E.}~\bibnamefont {Knill}}, \bibinfo {author} {\bibfnamefont {R.~W.}\
  \bibnamefont {Simmonds}}, \bibinfo {author} {\bibfnamefont {J.}~\bibnamefont
  {Aumentado}},\ and\ \bibinfo {author} {\bibfnamefont {J.~D.}\ \bibnamefont
  {Teufel}},\ }\bibfield  {title} {\bibinfo {title} {Direct observation of
  deterministic macroscopic entanglement},\ }\href
  {https://doi.org/10.1126/science.abf2998} {\bibfield  {journal} {\bibinfo
  {journal} {Science}\ }\textbf {\bibinfo {volume} {372}},\ \bibinfo {pages}
  {622} (\bibinfo {year} {2021})}\BibitemShut {NoStop}%
\bibitem [{\citenamefont {Teo}\ \emph {et~al.}(2017)\citenamefont {Teo},
  \citenamefont {M\"uller}, \citenamefont {Jeong}, \citenamefont {Hradil},
  \citenamefont {\ifmmode \check{R}\else \v{R}\fi{}eh\'a\ifmmode~\check{c}\else
  \v{c}\fi{}ek},\ and\ \citenamefont {S\'anchez-Soto}}]{PhysRevA.95.042322}%
  \BibitemOpen
  \bibfield  {author} {\bibinfo {author} {\bibfnamefont {Y.~S.}\ \bibnamefont
  {Teo}}, \bibinfo {author} {\bibfnamefont {C.~R.}\ \bibnamefont {M\"uller}},
  \bibinfo {author} {\bibfnamefont {H.}~\bibnamefont {Jeong}}, \bibinfo
  {author} {\bibfnamefont {Z.}~\bibnamefont {Hradil}}, \bibinfo {author}
  {\bibfnamefont {J.}~\bibnamefont {\ifmmode \check{R}\else
  \v{R}\fi{}eh\'a\ifmmode~\check{c}\else \v{c}\fi{}ek}},\ and\ \bibinfo
  {author} {\bibfnamefont {L.~L.}\ \bibnamefont {S\'anchez-Soto}},\ }\bibfield
  {title} {\bibinfo {title} {Superiority of heterodyning over homodyning: An
  assessment with quadrature moments},\ }\href
  {https://doi.org/10.1103/PhysRevA.95.042322} {\bibfield  {journal} {\bibinfo
  {journal} {Phys. Rev. A}\ }\textbf {\bibinfo {volume} {95}},\ \bibinfo
  {pages} {042322} (\bibinfo {year} {2017})}\BibitemShut {NoStop}%
\bibitem [{\citenamefont {Collett}\ and\ \citenamefont
  {Gardiner}(1984)}]{PhysRevA.30.1386}%
  \BibitemOpen
  \bibfield  {author} {\bibinfo {author} {\bibfnamefont {M.~J.}\ \bibnamefont
  {Collett}}\ and\ \bibinfo {author} {\bibfnamefont {C.~W.}\ \bibnamefont
  {Gardiner}},\ }\bibfield  {title} {\bibinfo {title} {Squeezing of intracavity
  and traveling-wave light fields produced in parametric amplification},\
  }\href {https://doi.org/10.1103/PhysRevA.30.1386} {\bibfield  {journal}
  {\bibinfo  {journal} {Phys. Rev. A}\ }\textbf {\bibinfo {volume} {30}},\
  \bibinfo {pages} {1386} (\bibinfo {year} {1984})}\BibitemShut {NoStop}%
\bibitem [{\citenamefont {Righini}\ \emph {et~al.}(2011)\citenamefont
  {Righini}, \citenamefont {Dumeige}, \citenamefont {F{\'e}ron}, \citenamefont
  {Ferrari}, \citenamefont {Nunzi~Conti}, \citenamefont {Ristic},\ and\
  \citenamefont {Soria}}]{Righini2011}%
  \BibitemOpen
  \bibfield  {author} {\bibinfo {author} {\bibfnamefont {G.~C.}\ \bibnamefont
  {Righini}}, \bibinfo {author} {\bibfnamefont {Y.}~\bibnamefont {Dumeige}},
  \bibinfo {author} {\bibfnamefont {P.}~\bibnamefont {F{\'e}ron}}, \bibinfo
  {author} {\bibfnamefont {M.}~\bibnamefont {Ferrari}}, \bibinfo {author}
  {\bibfnamefont {G.}~\bibnamefont {Nunzi~Conti}}, \bibinfo {author}
  {\bibfnamefont {D.}~\bibnamefont {Ristic}},\ and\ \bibinfo {author}
  {\bibfnamefont {S.}~\bibnamefont {Soria}},\ }\bibfield  {title} {\bibinfo
  {title} {Whispering gallery mode microresonators: Fundamentals and
  applications},\ }\href {https://doi.org/10.1393/ncr/i2011-10067-2} {\bibfield
   {journal} {\bibinfo  {journal} {Riv. Nuovo Cimento}\ }\textbf {\bibinfo
  {volume} {34}},\ \bibinfo {pages} {435} (\bibinfo {year} {2011})}\BibitemShut
  {NoStop}%
\bibitem [{\citenamefont {Schulman}\ \emph {et~al.}(2017)\citenamefont
  {Schulman}, \citenamefont {Wolski}, \citenamefont {Dhariwal}, \citenamefont
  {Radford},\ and\ \citenamefont {Klimov}}]{PPO}%
  \BibitemOpen
  \bibfield  {author} {\bibinfo {author} {\bibfnamefont {J.}~\bibnamefont
  {Schulman}}, \bibinfo {author} {\bibfnamefont {F.}~\bibnamefont {Wolski}},
  \bibinfo {author} {\bibfnamefont {P.}~\bibnamefont {Dhariwal}}, \bibinfo
  {author} {\bibfnamefont {A.}~\bibnamefont {Radford}},\ and\ \bibinfo {author}
  {\bibfnamefont {O.}~\bibnamefont {Klimov}},\ }\href
  {https://arxiv.org/abs/1707.06347} {\bibinfo {title} {Proximal policy
  optimization algorithms}} (\bibinfo {year} {2017}),\ \Eprint
  {https://arxiv.org/abs/1707.06347} {arxiv:1707.06347 [cs.LG]} \BibitemShut
  {NoStop}%
\end{thebibliography}%

\end{document}